\documentclass[10pt]{article}
\usepackage{amsmath}
\usepackage{graphicx}
\usepackage{color}
\usepackage{orcidlink}
\usepackage{hyperref}
\usepackage{multirow}
\usepackage{rotating}
\usepackage{multicol}
\hypersetup{colorlinks=true, linkcolor=blue, citecolor=blue, urlcolor=blue}
\usepackage{caption}
\usepackage{subcaption}
\usepackage{setspace}
\RequirePackage[numbers,sort&compress]{natbib}
\begin{document}
\baselineskip=15pt

\begin{center}
\LARGE{Noncommutative Geometry Inspired AdS Black Hole with a Cloud of Strings Surrounded by Quintessence-like fluid }
\par\end{center}

\vspace{0.3cm}

\begin{center}
{\bf Faizuddin Ahmed\orcidlink{0000-0003-2196-9622}}\footnote{\bf faizuddinahmed15@gmail.com}\\
{\tt Department of Physics, Royal Global University, Guwahati, 781035, Assam, India}\\
\vspace{0.1cm}
{\bf Allan. R. P. Moreira\orcidlink{0000-0002-6535-493X}}\footnote{\bf allan.moreira@fisica.ufc.br (Corresp. author)}\\ 
{\tt Secretaria da Educa\c{c}\~{a}o do Cear\'{a} (SEDUC), Coordenadoria Regional de Desenvolvimento da Educa\c{c}\~{a}o (CREDE 9),  Horizonte, Cear\'{a}, 62880-384, Brazil}\\
\vspace{0.1cm}
{\bf Abdelmalek Bouzenada\orcidlink{0000-0002-3363-980X}}\footnote{\bf abdelmalekbouzenada@gmail.com}\\ 
{\tt Laboratory of Theoretical and Applied Physics, Echahid Cheikh Larbi Tebessi University 12001, Algeria}\\
\vspace{0.1cm}

\end{center}

\vspace{0.3cm}

\begin{abstract}
We investigate the geometric and physical properties of an anti-de Sitter (AdS) black hole space-time coupled by a cloud of strings and surrounded by a quintessence-like fluid, all within the framework of non-commutative (NC) geometry. From the perspective of geometrical optics, we analyze the behavior of null geodesics, focusing on key optical features such as the effective potential, the structure and radius of the photon sphere, light deflection angles, photon trajectories, and the resulting black hole (BH) shadow. Our findings show that the combined effects of the string cloud and quintessence-like fluid significantly modify photon dynamics and optical observables, leading to notable deviations from standard BH scenarios in NC geometry background. We also examine time-like geodesics, with particular emphasis on the innermost stable circular orbits (ISCOs). The results demonstrate that the presence of geometric matter components alters the ISCO radius compared to conventional solutions. In addition, we explore the thermodynamic behavior of the BH, deriving expressions for the Hawking temperature, entropy, Gibbs free energy, and specific heat capacity. The influence of the string cloud and quintessence-like fluid introduces substantial modifications to the thermodynamic profile, including shifts in phase transition points and changes to stability conditions under NC geometric effects. Furthermore, we study the dynamics of massless scalar field perturbations in this modified background by formulating the Klein-Gordon equation and reducing it to a Schrödinger-like form via separation of variables. The resulting effective potential is analyzed in detail, highlighting the significant role of both the string cloud and the quintessence-like fluid in shaping the perturbative landscape within the NC geometry setting. 
\end{abstract}

\vspace{0.3cm}


\section{Introduction} \label{intro}

General Relativity (GR) \cite{AZ1} is a cornerstone of modern theoretical physics, providing a robust mathematical framework for describing gravitational interactions. It has been extensively tested and successfully explains a wide range of astrophysical phenomena \cite{AZ2}, including the deflection of light by massive bodies, gravitational wave emission and propagation \cite{AZ3}, and the intricate dynamics of black holes (BHs) \cite{AZ4}. GR also predicts the formation of spacetime singularities regions of infinite curvature at the centers of BHs and at the universe's origin.

Despite its successes, GR faces major challenges. It is fundamentally classical and does not incorporate quantum mechanics (QM), which governs physics at microscopic scales. This incompatibility leaves the quest for a consistent theory of quantum gravity (QG) unresolved. Additionally, GR does not account for dark matter \cite{AZ5} or dark energy \cite{AZ6}, highlighting the need for extended gravitational theories. In response, various modified gravity models have been proposed, often inspired by string theory and high-energy physics (HEP), where higher-dimensional curvature terms are added to the Einstein-Hilbert (EH) action to explore extreme energy regimes. These extensions help probe the limits of GR and its breakdown in strong-field regions. Alternative approaches like Loop Quantum Gravity (LQG) and string theory introduce quantum corrections to classical geometries, especially near BH horizons and singularities \cite{AZ7,AZ8,AZ9,AZ10}. These frameworks aim to resolve singularities into regular, finite structures and offer insights into the quantum nature of spacetime \cite{AZ11,AZ12,AZ13,AZ14,AZ15,AZ16}.

The advent of advanced astronomical observatories has enabled direct tests of General Relativity (GR) in previously inaccessible regimes. A landmark achievement is the imaging of supermassive black holes, notably M87* \cite{AZ17} and Sgr A* \cite{AZ18}, by the Event Horizon Telescope (EHT). These observations reveal BH shadows-shaped by photon trajectories near the event horizon—which are strongly influenced by the space-time geometry. The close match between observed shadows and predictions from the Kerr solution in GR reinforces the theory's validity in the strong-field regime \cite{AZ19,AZ20}. 
However, quantum gravitational effects-though negligible under typical conditions—may introduce subtle deviations in observables such as shadow structure, light deflection, or gravitational lensing. These deviations could offer insight into quantum-corrected black hole models \cite{AZ21,AZ22,AZ23,AZ24,AZ25}. Continued theoretical advances, combined with high-precision observations, may ultimately lead to a unified theory that reconciles GR with quantum mechanics.

Non-commutative (NC) black holes (BHs) have advanced the search for quantum gravity by introducing a minimal length scale that can potentially resolve spacetime singularities \cite{NC1,NC2,NC3}. This concept, supported by string theory and spacetime quantization, replaces classical point-like masses with smeared matter distributions, ensuring regular geometries \cite{NC4,NC5}. NC geometry modifies the energy-momentum tensor while preserving the Einstein tensor, thus retaining the geometric framework of general relativity with quantum corrections \cite{NC7}.

A foundational result by Nicolini {\it et al.} established the NC Schwarzschild solution, which eliminates the central singularity and asymptotically approaches the classical Schwarzschild metric \cite{NC7}. Susskind emphasized string effects in BH complementarity \cite{NC6}, and Seiberg {\it et al.} demonstrated how NC space-time emerges from string theory in the presence of background gauge fields \cite{NC8}. These frameworks have been extended to charged \cite{NC11}, rotating \cite{NC13}, and higher-dimensional BHs \cite{NC12,NC14,NC15}, including charged configurations in extra dimensions \cite{NC16,NC17}. Thermodynamic properties-such as Hawking radiation, entropy corrections, and phase transitions have been extensively studied in NC BHs \cite{NC9,NC18,NC19,NC20,NC21}. NC effects also alter wave dynamics, affecting absorption and scattering cross sections \cite{NC22,NC23,NC24,NC25}. Since the 1970s, scalar and fermionic wave scattering studies have contributed to BH identification \cite{NC26,NC27,NC28,NC29,NC30,NC31}, with the partial wave method proving particularly effective \cite{NC32,NC33,NC34,NC35,NC36}. Recent investigations explored scalar wave scattering in NC BHs using Gaussian and Lorentzian smearing \cite{NC37}, effects of magnetic fields \cite{NC38}, and extended analyses to Reissner-Nordström \cite{NC39}, Bardeen \cite{NC40}, and BTZ BHs \cite{NC41}. Higher-dimensional and string-theoretic NC BHs reveal even richer scattering phenomena \cite{NC42}.

The primary objective of this work is to investigate how noncommutative geometry, combined with external matter sources-a cloud of strings and a quintessence-like fluid modifies the physical and geometric properties of AdS BH solution. We systematically analyze the implications of these corrections on photon dynamics, space-time structure, ISCO radius, thermodynamic behavior, and scalar field perturbations, contrasting our findings with classical black hole predictions. These modifications introduce non-trivial deviations in geodesic motion, thermodynamic stability, and perturbative responses, potentially yielding observable signatures in black hole phenomena. First, we examine the motion of massless particles by studying null geodesics, focusing on how the effective potential, circular photon orbits, photon sphere, light bending, and black hole shadow are influenced by the interplay of noncommutative geometry, string clouds, and quintessence. Next, we derive key thermodynamic quantities-including Hawking temperature, entropy, Gibbs free energy, and specific heat capacity-to assess how these additional structures affect phase transitions and thermal stability. Finally, we study scalar field perturbations in this modified background, computing the effective potential to evaluate the stability of the black hole solution under such deformations. By integrating these analyses, our work provides a comprehensive framework for understanding how quantum geometric effects and extended matter fields alter classical black hole physics in AdS spacetimes, offering insights into potential observable consequences in high-energy and gravitational-wave astronomy.

This paper is organized as follows: In Section \ref{sec:2}, we examine the geodesic motion of photon particles in the modified black hole spacetime, highlighting the effects of string cloud and quintessence field contributions on photon dynamics within the noncommutative geometry. In Section \ref{sec:3}, we investigate the thermodynamic properties of the BH solution, analyzing quantities such as the Hawking temperature, entropy, and specific heat, and discussing how these are influenced by the underlying physical modifications. In Section \ref{sec:4}, we study scalar field perturbations in the BH background by deriving and solving the massless Klein-Gordon equation. We analyze the resulting effective potential and its implications for stability. Finally, in Section \ref{sec:5}, we summarize the main results and discuss their broader implications for observational astrophysics and theoretical developments in BH physics. 

\section{NC geometry inspired AdS BH coupled with CS surrounded by QF }\label{sec:2}

The metric of a Schwarzschild-AdS BH within the framework of noncommutative geometry has been studied and is provided in \cite{RBW}. This space-time is given by
\begin{eqnarray}
    ds^2=-\left(1-\frac{2\,M}{r}+\frac{\lambda\,M}{r^2}-\frac{\Lambda}{3}\,r^2\right)\,dt^2+\left(1-\frac{2\,M}{r}+\frac{a\,M}{r^2}-\frac{\Lambda}{3}\,r^2\right)^{-1}\,dr^2 +r^2\,(d\theta^2+\sin^2\theta\,d\phi^2),\label{aa2}
\end{eqnarray}
where \(\lambda\) is related with non-commutative parameter $\Theta$ whose dimension is squared length given by $\lambda=8\,\sqrt{\frac{\Theta}{\pi}}$. Here, \( M \) denotes the BH mass, and \( \Lambda \) is the cosmological constant.

The charged static BH with cosmological constant and surrounded by a cloud of strings (CS) and quintessence field (QF) was discussed in \cite{JMT}. This charged space-time with CS and QF is given by
\begin{eqnarray}
    ds^2=-\left(1-\alpha-\frac{2\,M}{r}+\frac{Q^2}{r^2}-\frac{\mathrm{N}}{r^{3\,w+1}}-\frac{\Lambda}{3}\,r^2\right)\,dt^2+\left(1-\frac{2\,M}{r}+\frac{Q^2}{r^2}-\frac{\mathrm{N}}{r^{3\,w+1}}-\frac{\Lambda}{3}\,r^2\right)^{-1}\,dr^2 +r^2\,(d\theta^2+\sin^2\theta\,d\phi^2),\label{aa2aa}
\end{eqnarray}
where $(\mathrm{N}, w)$ are the quintessence field parameters.

Therefore, we introduce the line element of an AdS BH surrounded by a cloud of strings and quintessence field in noncommutative geometric background, given by 
\begin{equation}
     ds^2=-\mathcal{F} (r)\,dt^2+\frac{dr^2}{\mathcal{F} (r)}+r^2\,(d\theta^2+r^2\,\sin^2 \theta\,d\phi^2),\label{aa3}
\end{equation}
where the metric function $\mathcal{F}(r)$ is given by
\begin{equation}
     \mathcal{F} (r)=1-\alpha - \frac{2\,M}{r}+\frac{\lambda\,M}{r^2}-\frac{\mathrm{N}}{r^{3\,w+1}} - \frac{\Lambda}{3}\,r^2\quad\quad (-1 <w < -1/3).\label{aa4}
\end{equation}
The range of the coordinates are as follows:
\begin{equation}
     -\infty < t < +\infty,\quad\quad r\geq 0,\quad\quad 0 \leq \theta \leq \pi,\quad\quad 0 \leq \phi < 2\,\pi.\label{aa5}
\end{equation}

\begin{figure}[ht!]
\centering
\includegraphics[height=3.5cm,width=4.7cm]{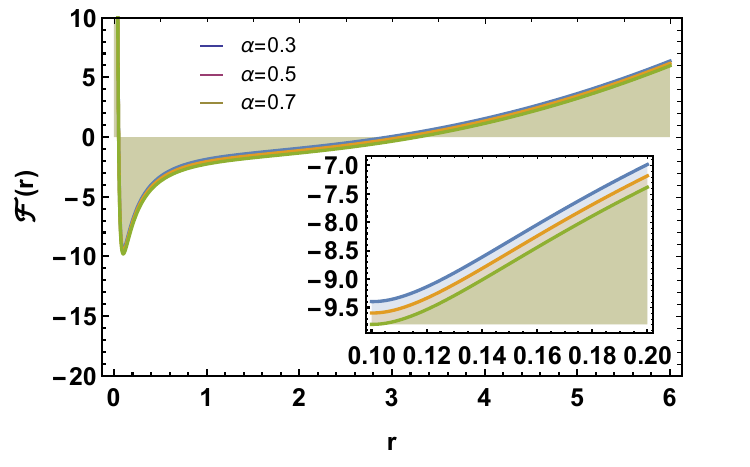}
\includegraphics[height=3.5cm,width=4.7cm]{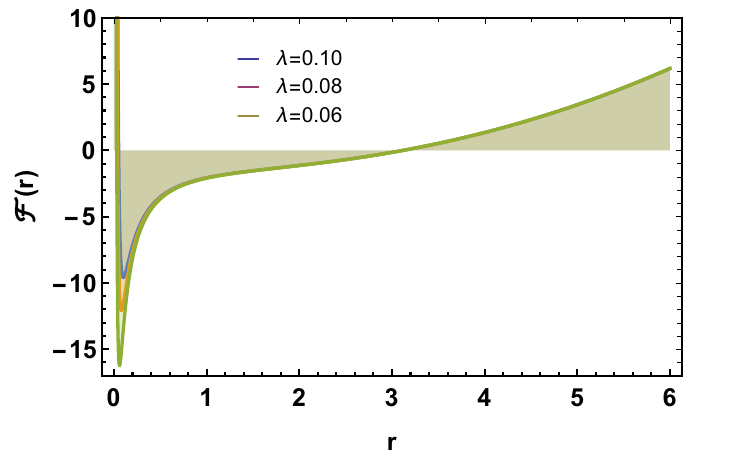}\\
\includegraphics[height=3.5cm,width=4.7cm]{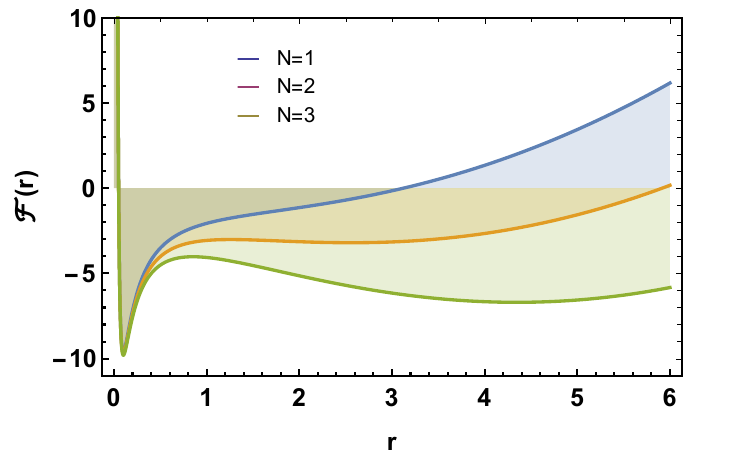}
\includegraphics[height=3.5cm,width=4.7cm]{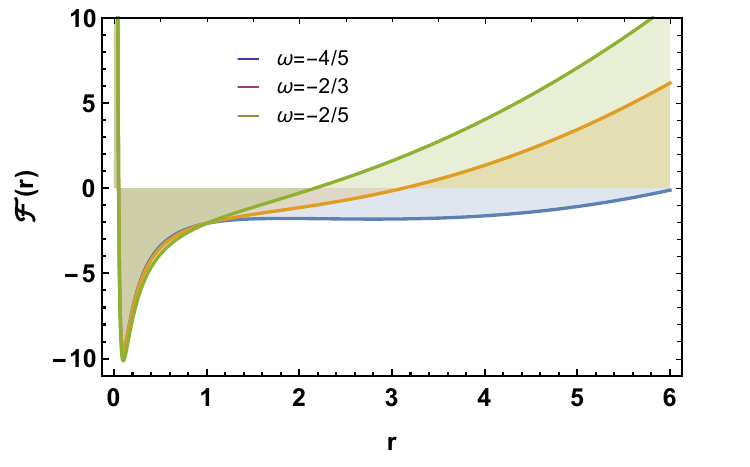}\\
(a) \hspace{4cm} (b) \hspace{4cm} (c) \hspace{4cm} (d)
\caption{\footnotesize Behavior of the metric coefficient $\mathcal{F} (r)$ vs the radial coordinate $r$ for different values of the geometrical parameters. (a) $\Lambda=-1$, $\lambda=0.1$, $\omega=-2/3$ and $N=1$. (b) $\Lambda=-1$, $\alpha=0.3$, $\omega=-2/3$ and $N=1$. (c) $\Lambda=-1$, $\alpha=0.3$, $\lambda=0.1$ and $\omega=-2/3$. (d) $\Lambda=-1$, $\alpha=0.3$, $\lambda=0.1$ and $N=1$.}
\label{fig3}
\end{figure}

\begin{figure}[ht!]
\centering
\includegraphics[height=4cm,width=5cm]{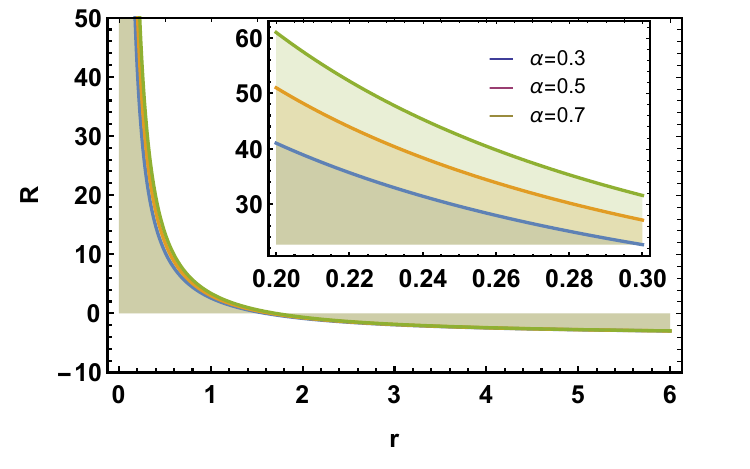}
\includegraphics[height=4cm,width=5cm]{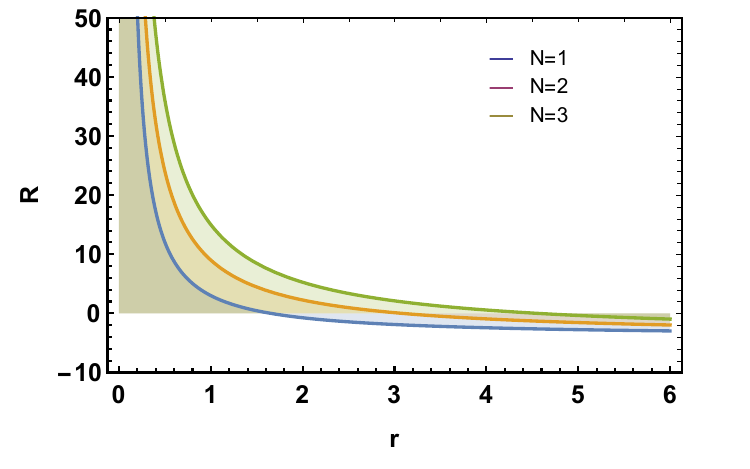}
\includegraphics[height=4cm,width=5cm]{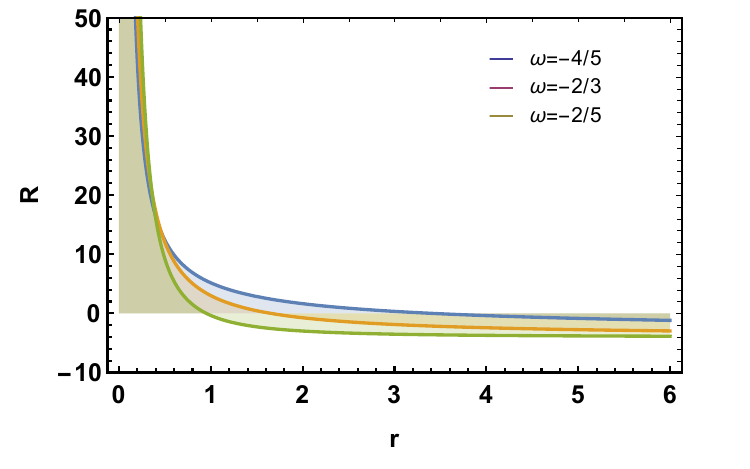}\\
(a) \hspace{5cm}  (b) \hspace{5cm} (c)
\caption{\footnotesize Behavior of the curvature scalar $R$ vs the radial coordinate $r$ for different values of the geometrical parameters. (a) $\Lambda=-1$, $\omega=-2/3$ and $N=1$. (b) $\Lambda=-1$, $\alpha=0.3$, and $\omega=-2/3$ (c)  $\Lambda=-1$, $\alpha=0.3$ and $N=1$.}
\label{fig2.1}
\hfill\\
\includegraphics[height=4cm,width=5cm]{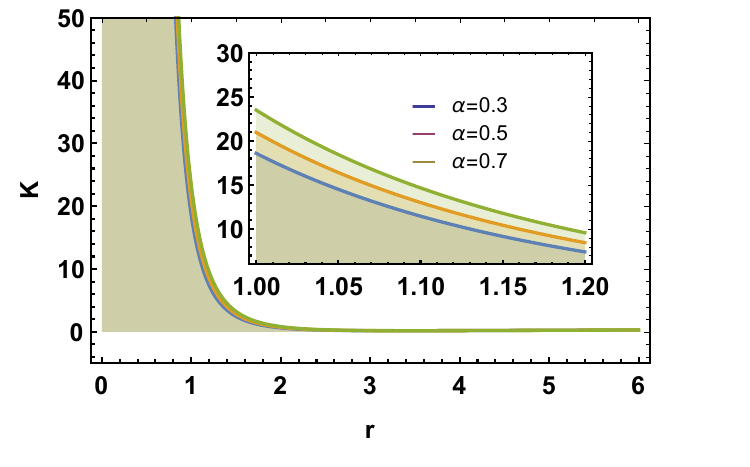}
\includegraphics[height=4cm,width=5cm]{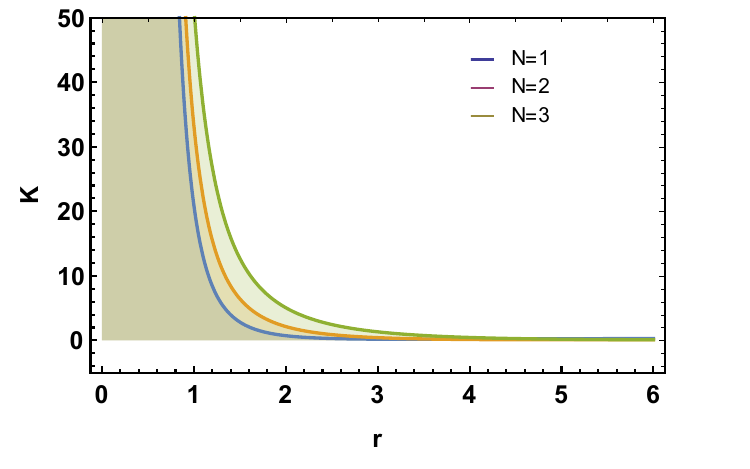}
\includegraphics[height=4cm,width=5cm]{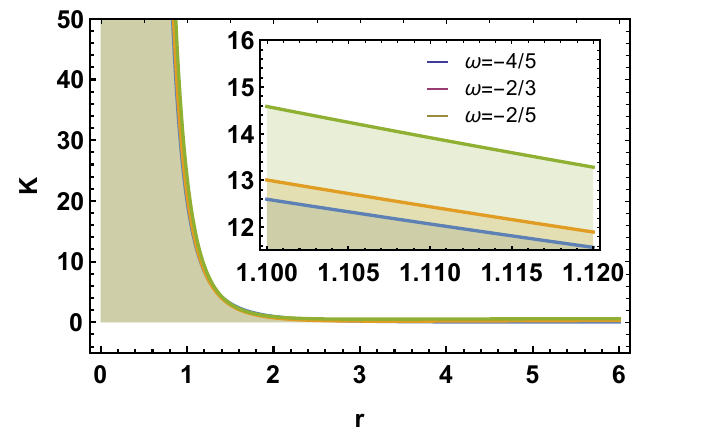}\\
(a) \hspace{5cm}  (b) \hspace{5cm} (c)
\caption{\footnotesize Behavior of the Kretschmann scalar $K$ vs the radial coordinate $r$ for different values of the geometrical parameters.(a) $\Lambda=-1$, $\lambda=0.1$, $\omega=-2/3$ and $N=1$. (b) $\Lambda=-1$, $\alpha=0.3$, $\lambda=0.1$ and $\omega=-2/3$. (c) $\Lambda=-1$, $\alpha=0.3$, $\lambda=0.1$ and $N=1$.}
\label{fig3.1}
\end{figure}

As we can see, Fig.~\ref{fig3} displays the radial dependence of the metric function $\mathcal{F}(r)$, which governs the causal structure of the black hole space-time and plays a fundamental role in thermodynamic analysis. In  Fig.~\ref{fig3} (a), varying the cloud of strings parameter $\alpha$ reveals that increasing $\alpha$ suppresses the depth of the potential well near the origin and slightly shifts the location of the event horizon, as shown in the inset. This behavior reflects the tension induced by the string cloud, which effectively screens the gravitational pull.  Fig.~\ref{fig3} (b) focuses on the influence of the noncommutative parameter $\lambda$, showing that higher values of $\lambda$ smooth out the metric profile near $r \rightarrow 0$, confirming the expected regularizing effect of the Lorentzian noncommutative background. In  Fig.~\ref{fig3} (c), the effect of the quintessence normalization $N$ is shown to induce deeper gravitational wells as $N$ increases, which correspondingly shifts the horizon to smaller radii. Finally,  Fig.~\ref{fig3} (d) investigates variations in the equation-of-state parameter $\omega$, revealing that more negative values of $\omega$ contribute more significantly to the inward curvature of $\mathcal{F}(r)$, deepening the potential and shifting the horizon location. These modifications to $\mathcal{F}(r)$ impact the thermodynamic quantities—especially the Hawking temperature and Gibbs free energy—by altering both the surface gravity and the mass function $M(r_+)$, as shown in Eqs.~(\ref{ff5}) and (\ref{ff11}). Overall, the interplay between $\alpha$, $\lambda$, $N$, and $\omega$ not only affects the causal structure but also dictates the thermodynamic stability and phase transitions of the noncommutative AdS black hole.

Figure~\ref{fig2.1} illustrates the behavior of the Ricci scalar curvature $R$ as a function of the radial coordinate $r$ for different configurations of the model parameters. In Fig.\ref{fig2.1} (a), we explore the effect of the string cloud parameter $\alpha$, where increasing values of $\alpha$ lead to a higher curvature near the origin, indicating a more pronounced gravitational field in the strong-field regime. Fig.\ref{fig2.1} (b) analyzes the impact of the quintessence normalization parameter $N$, showing that larger $N$ intensifies the curvature at small $r$, consistent with the behavior of exotic matter contributions. In Fig.\ref{fig2.1} (c), varying the equation-of-state parameter $\omega$ within the allowed range $(-1 < \omega < -1/3)$ reveals that more negative values of $\omega$ yield stronger curvature near the core, in agreement with standard quintessence models.

Figure~\ref{fig3.1} presents the behavior of the Kretschmann scalar $K = R_{\mu\nu\rho\sigma}R^{\mu\nu\rho\sigma}$ as a function of the radial coordinate $r$, which serves as a diagnostic tool for identifying curvature singularities in the space-time. In Fig.\ref{fig3.1}(a), we analyze the effect of the string cloud parameter $\alpha$, showing that higher values of $\alpha$ lead to a stronger divergence of $K$ near the origin, emphasizing the role of string tension in amplifying curvature effects. The inset highlights this sensitivity in the near-horizon region. In Fig.\ref{fig3.1} (b), varying the quintessence normalization $N$ demonstrates that an increase in $N$ leads to more pronounced curvature at small $r$, consistent with the enhanced gravitational contribution from the quintessence field. Finally, Fig.\ref{fig3.1} (c) explores the impact of the equation-of-state parameter $\omega$, revealing that more negative values of $\omega$ result in steeper gradients of $K$ close to the core. Notably, across all configurations, the Kretschmann scalar diverges as $r \to 0$, signaling the presence of a curvature singularity.

\section{Geometrical Properties of BH}

In this section, we investigate the geodesic motion of test particles in the spacetime of the selected black hole (BH) solution. Our primary objective is to understand how the underlying geometrical parameters of the spacetime influence key features such as particle trajectories, the conditions for circular orbits, the location of the photon sphere, and the resulting shadow size. The study of geodesics is a fundamental tool for probing the geometrical and physical properties of black hole spacetimes. It provides insight into both the gravitational influence of the black hole and the observational signatures that could potentially distinguish between different BH models.

Given that the spacetime under consideration is static and spherically symmetric, the analysis of geodesics can be significantly simplified by restricting the motion of test particles to the equatorial plane, defined by  \( \theta = \pi/2 \). This restriction does not lead to any loss of generality due to the symmetry of the system. To examine particle motion in curved spacetime, we employ the Lagrangian formalism, which allows us to derive the equations of motion from a variational principle. For the given metric, the Lagrangian density function $\mathcal{L}=\frac{1}{2}\,g_{\mu\nu}\,\dot{x}^{\mu}\,\dot{x}^{\nu}$, where dot represents an ordinary derivative w. r. to an affine parameter along the geodesic and $g_{\mu\nu}$ denotes the components of the metric tensor. This formulation serves as the starting point for deriving the geodesic equations through the Euler–Lagrange equations. References relevant to this analysis include \cite{FA1,FA2,FA3,FA4,FA5,FA6,FA7,FA8,FA9,FA10,FA11,FA12}, which discuss similar methodologies applied to various modified gravity black hole space-times.

The Lagrangian density function $\mathcal{L}$ in the equatorial plane using metric (\ref{aa3}) is given by  
\begin{equation}\label{bb1}
    \mathcal{L}=\frac{1}{2}\,\Big[-\mathcal{F} (r)\,\dot{t}^2+\frac{\dot{r}^2}{\mathcal{F} (r)}+r^2\,\dot{\phi}^2\Big].
\end{equation}
Since $\mathcal{L}$ does not depends on the coordinates $t$ and $\phi$, the system possesses two conserved quantities due to the associate symmetries. There are two Killing vectors, namely, the temporal or time translational: $\eta_{(t)} \equiv \partial_{t}$ and angular one $\eta_{(\phi)} \equiv \partial_{\phi}$. The conserved quantities are given by
\begin{eqnarray}
    &&\dot{t}=\frac{\mathrm{E}}{\mathcal{F} (r)},\label{bb2}\\
    &&\dot{\phi}=\frac{\mathrm{L}}{\,r^2}.\label{bb3}
\end{eqnarray}
where $\mathrm{E}$ is the conserved energy and $\mathrm{L}$ is the conserved angular momentum.

Using (\ref{bb2})--(\ref{bb3}) into the equation (\ref{bb1}), we find geodesics equation for the radial coordinate $r$ as, 
\begin{eqnarray}
    \dot{r}^2+\mathcal{F}\,\frac{\mathrm{L}^2}{\,r^2}=\mathrm{E}^2,\label{bb4}
\end{eqnarray}
where the effective potential $V_\text{eff}$ is given by
\begin{eqnarray}
     V_\text{eff}(r)=\mathcal{F}\,\left(-\epsilon+\frac{\mathrm{L}^2}{\,r^2}\right)=\left(-\epsilon+\frac{\mathrm{L}^2}{\,r^2}\right)\,\left[1-\alpha - \frac{2\,M}{r}+\frac{\lambda\,M}{r^2}-\frac{\mathrm{N}}{r^{3\,w+1}} - \frac{\Lambda}{3}\,r^2\right]. \label{bb5}
\end{eqnarray}

From the above expression (\ref{bb5}), it is clear that the effective potential for null geodesics depends on various geometrical parameters. These include  string clouds characterized by the parameter \( \alpha \), quintessential field parameters \((\mathrm{N},w)\), non-commutative geometry effects characterized by the parameter \( \lambda \), the BH mass \( M \), the cosmological constant \( \Lambda \), and the angular momentum \( \mathrm{L} \).

\subsection{Null Geodesic}

We now examine the trajectory equations governing the motion of photons in the given gravitational field and analyze how the string clouds, the quintessence field, and the noncommutative geometry influences their trajectories. 

For null geodesics, the effective potential $V_\text{eff}$ is given by
\begin{eqnarray}
     V_\text{eff}(r)=\frac{\mathrm{L}^2}{r^2}\,\left[1-\alpha - \frac{2\,M}{r}+\frac{\lambda\,M}{r^2}-\frac{\mathrm{N}}{r^{3\,w+1}} - \frac{\Lambda}{3}\,r^2\right]. \label{bb55}
\end{eqnarray}

\begin{figure}[ht!]
    \centering
    \includegraphics[width=0.3\linewidth]{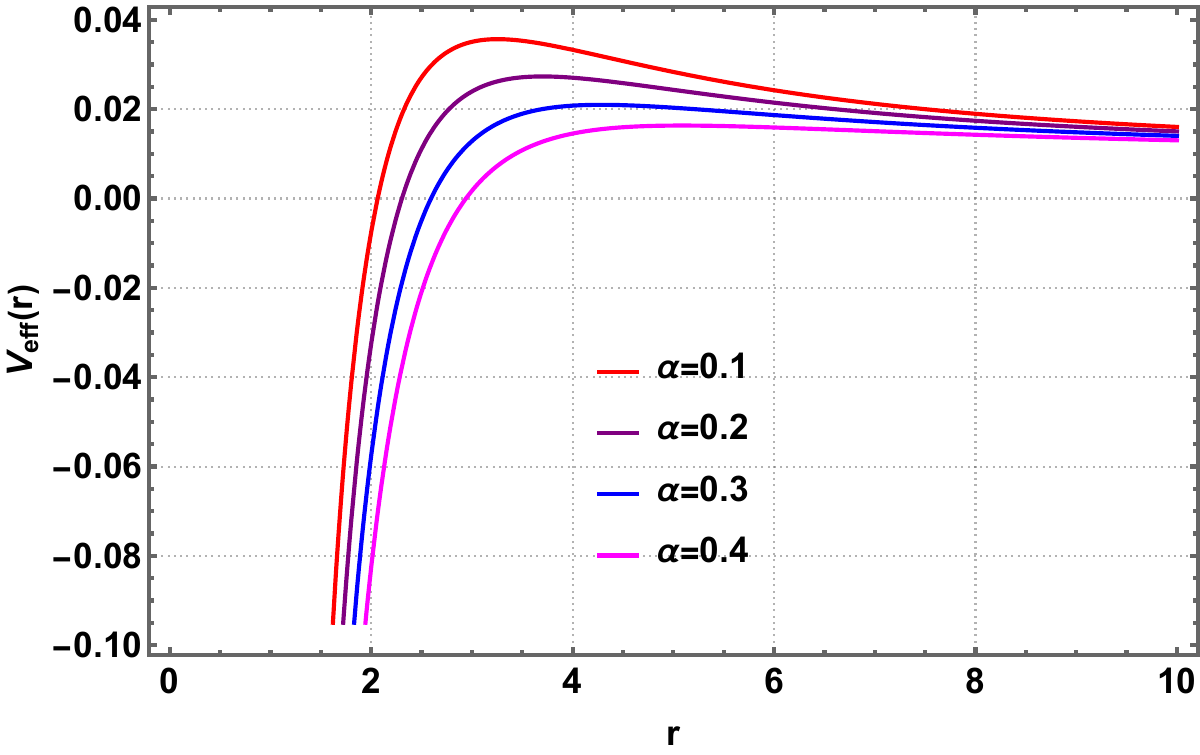}\quad\quad
    \includegraphics[width=0.3\linewidth]{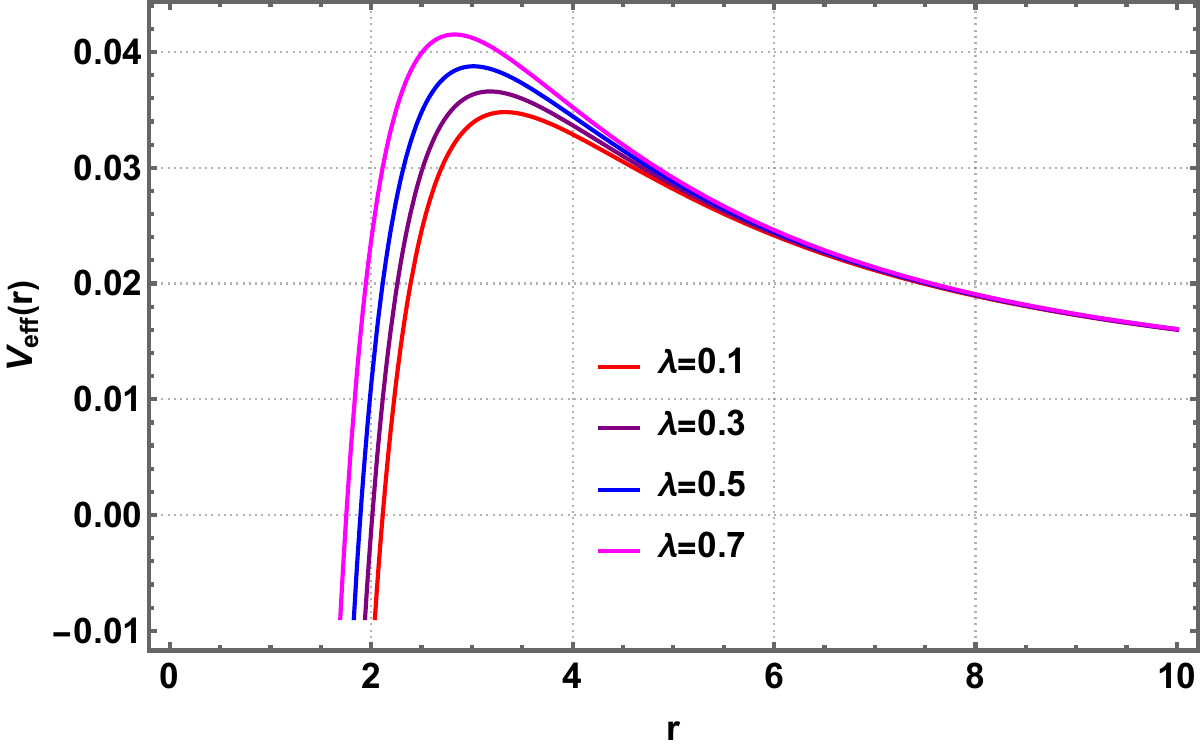}\quad\quad
    \includegraphics[width=0.3\linewidth]{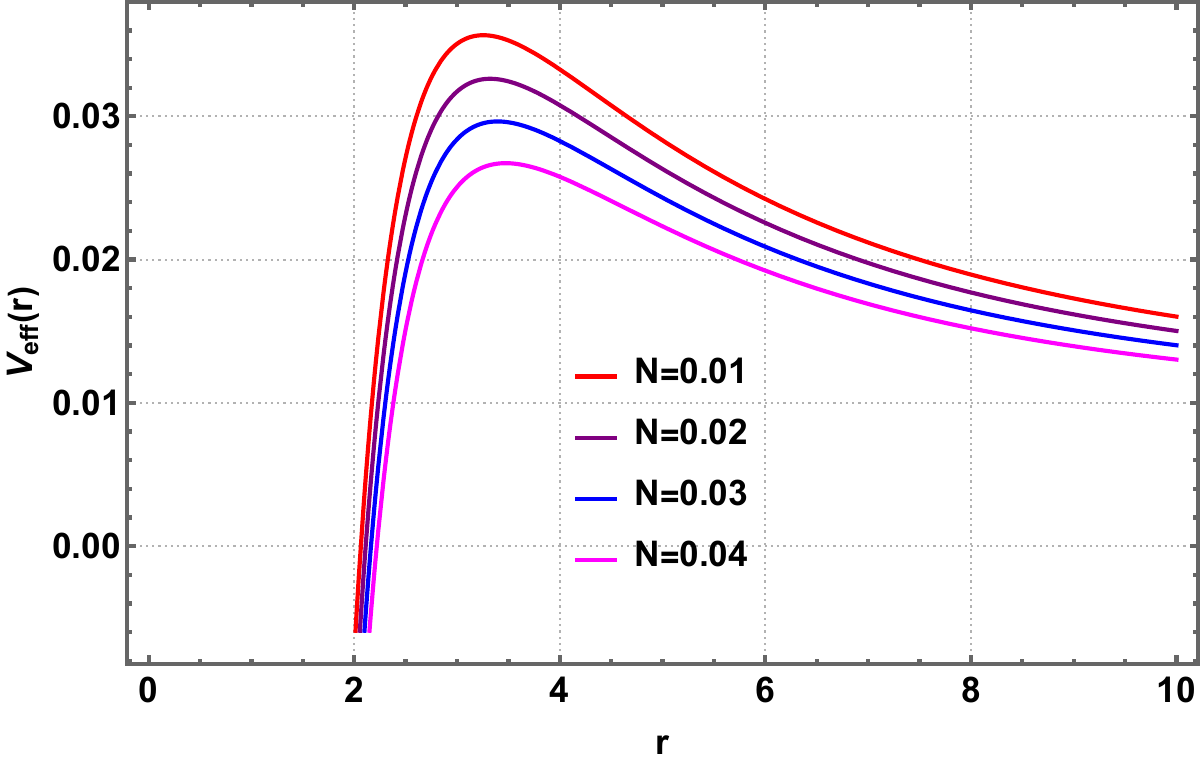}\\
    (a) $\lambda=0.2,\,\mathrm{N}=0.01$ \hspace{4cm} (b) $\alpha=0.1,\,\mathrm{N}=0.01$ \hspace{4cm} (c) $\lambda=0.2,\,\alpha=0.1$
    \caption{\footnotesize Behavior of the effective potential for the null geodesic by varying values of the string parameter \(\alpha\), NC parameter \(\lambda\), and the normalization constant $\mathrm{N}$. Here, we set $M=1=\mathrm{L}, w=-2/3$.}
    \label{fig:null-geodesic}
\end{figure}

In Figure \ref{fig:null-geodesic}, we present the effective potential for null geodesic motion as functions of $r$ for different values of the string parameter \(\alpha\), NC geometry parameter \(\lambda\), and the normalization constant \(\mathrm{N}\) associated with the quintessence field. Panels (a) and (c) show that the effective potential decreases as \(\alpha\) and \(\mathrm{N}\) increase, respectively, indicating that higher values of these parameters lower the potential barrier. Conversely, panel (b) demonstrates that the effective potential increases with increasing values of the NC geometry parameter \(\lambda\), suggesting a higher impact of this parameter on the  gravitational influence.

Using Eqs. (\ref{bb3}) and (\ref{bb4}) and finally (\ref{bb55}), we define the equation of photon orbit as follow:
\begin{eqnarray}
    \left(\frac{1}{r^2}\,\frac{dr}{d\phi}\right)^2=\frac{1}{\beta^2}+\frac{\Lambda}{3}-\frac{1-\alpha}{r^2}+\frac{2\,M}{r^3}-\frac{\lambda\,M}{r^4}+\frac{\mathrm{N}}{r^{3\,w+3}}. \label{bb6}
 \end{eqnarray}
where $\beta =\mathrm{L}/\mathrm{E}$ is the impact parameter of photons.
  
To simplify the analysis, we perform a transformation to a new variable via $u=1/r$. Substituting this into the Eq.  (\ref{bb6}), we obtain
\begin{eqnarray}
  \left(\frac{du}{d\phi}\right)^2+(1-\alpha)u^2=\frac{1}{\beta^2}+\frac{\Lambda}{3}+2\,M\,u^3-\lambda\,M\,u^4+\mathrm{N}\,u^{3\,w+3}.\label{bb7}
\end{eqnarray}
To analyze the photon trajectories (such as bending of light), it is customary to obtain a second-order differential equation. This is done by differentiating both sides of Eq.  (\ref{bb7}) w. r. t $\phi$ and after simplification, we get
\begin{eqnarray}
    \frac{d^2u}{d\phi^2}+(1-\alpha)\,u =3\,M\,u^2-2\,\lambda\,M\,u^3+\frac{3\,(w+1)}{2}\,\mathrm{N}\,u^{3\,w+2}.\label{bb8}  
\end{eqnarray}

\begin{figure}[ht!]
    \centering
    \includegraphics[width=0.25\linewidth]{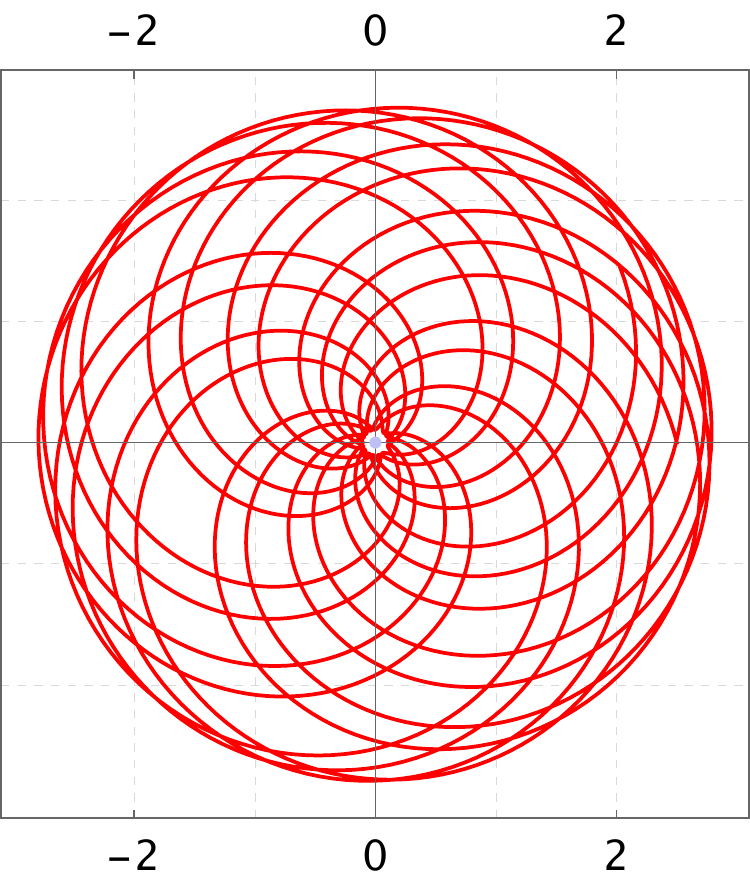}\quad\quad
    \includegraphics[width=0.25\linewidth]{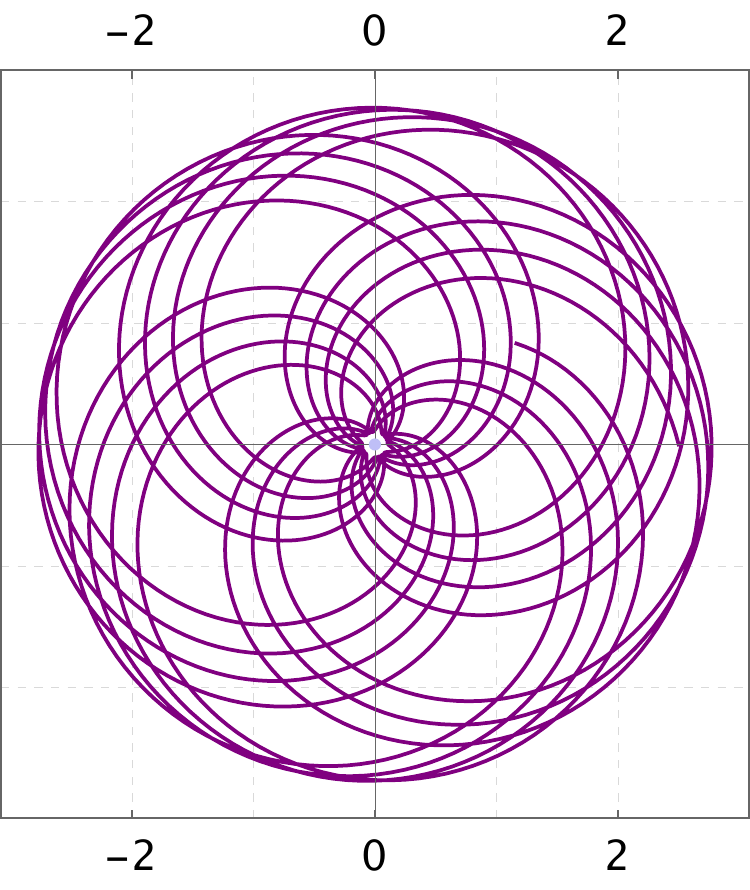}\quad\quad
    \includegraphics[width=0.25\linewidth]{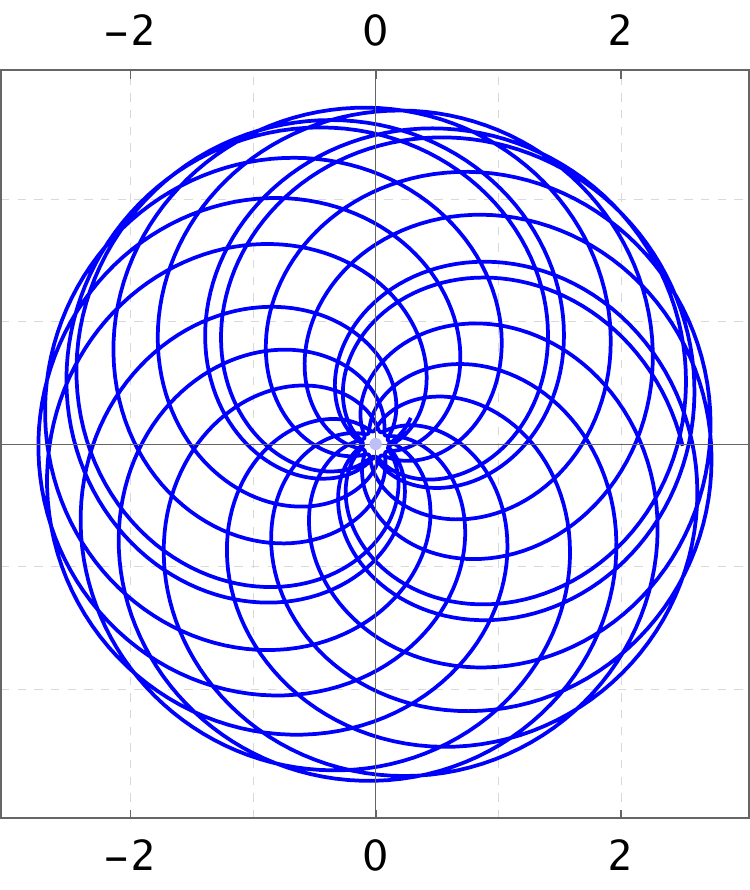}\\
    (a) $\alpha=0.10$ \hspace{4cm} (b) $\alpha=0.12$ \hspace{4cm} (c) $\alpha=0.14$\\
    \includegraphics[width=0.25\linewidth]{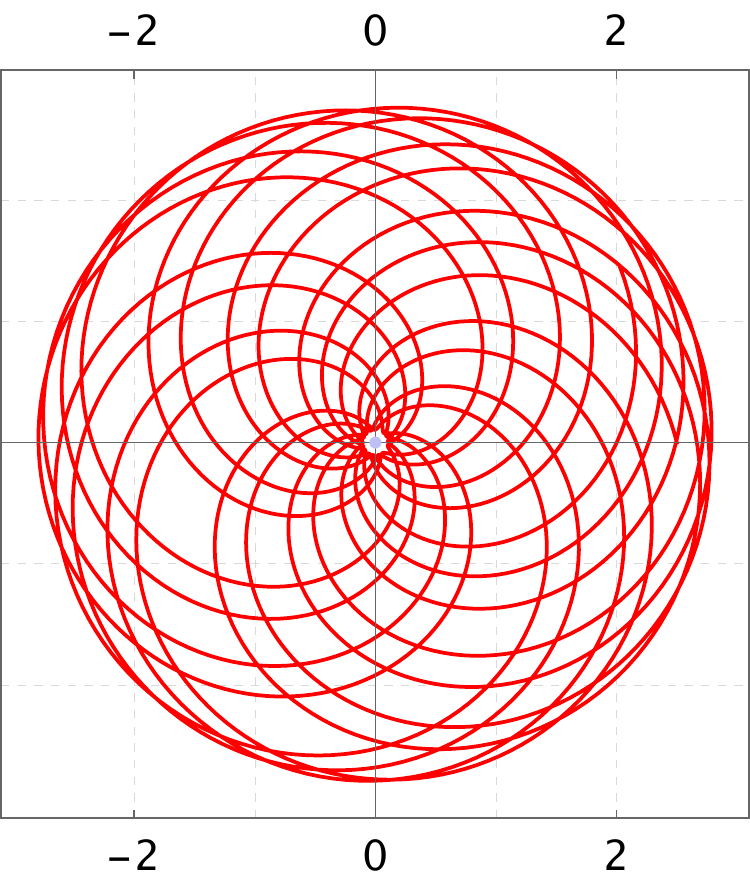}\quad\quad
    \includegraphics[width=0.25\linewidth]{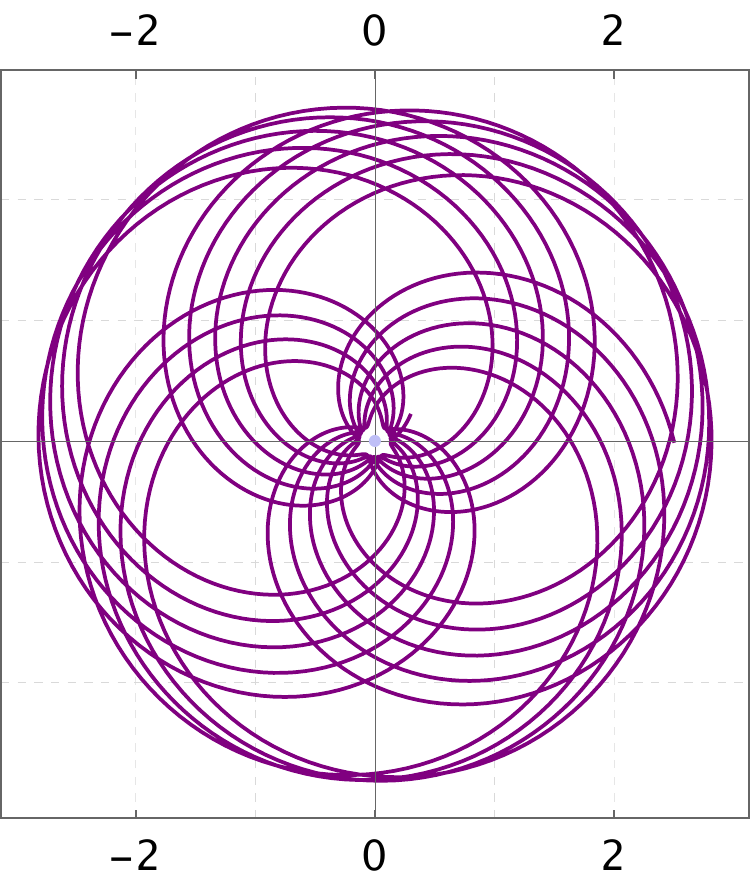}\quad\quad
    \includegraphics[width=0.25\linewidth]{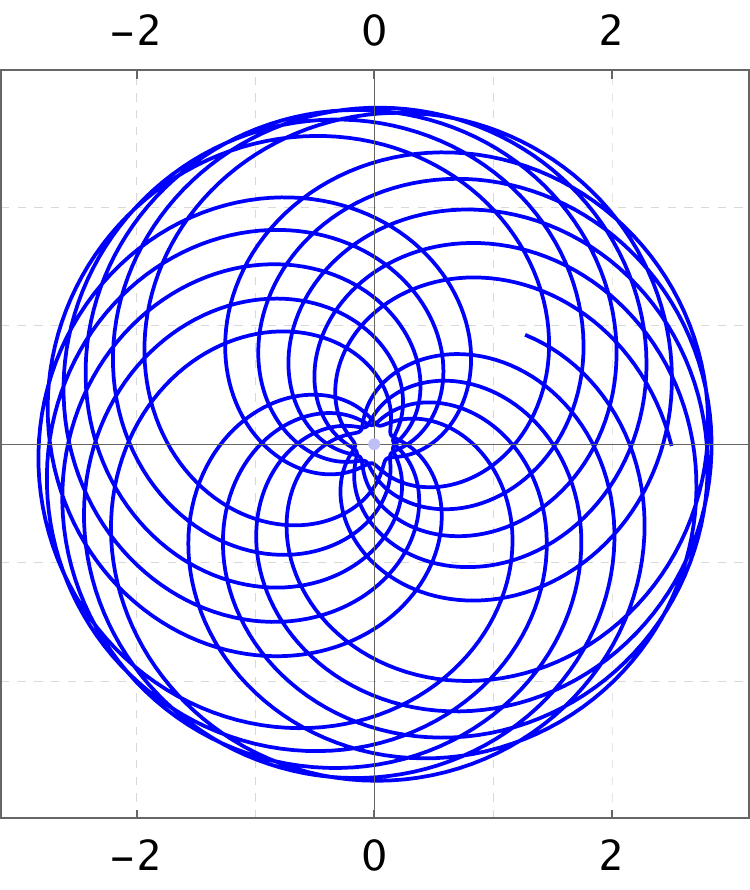}\\
    (d) $\lambda=0.20$ \hspace{4cm} (e) $\lambda=0.25$ \hspace{4cm} (f) $\lambda=0.30$\\
    \includegraphics[width=0.25\linewidth]{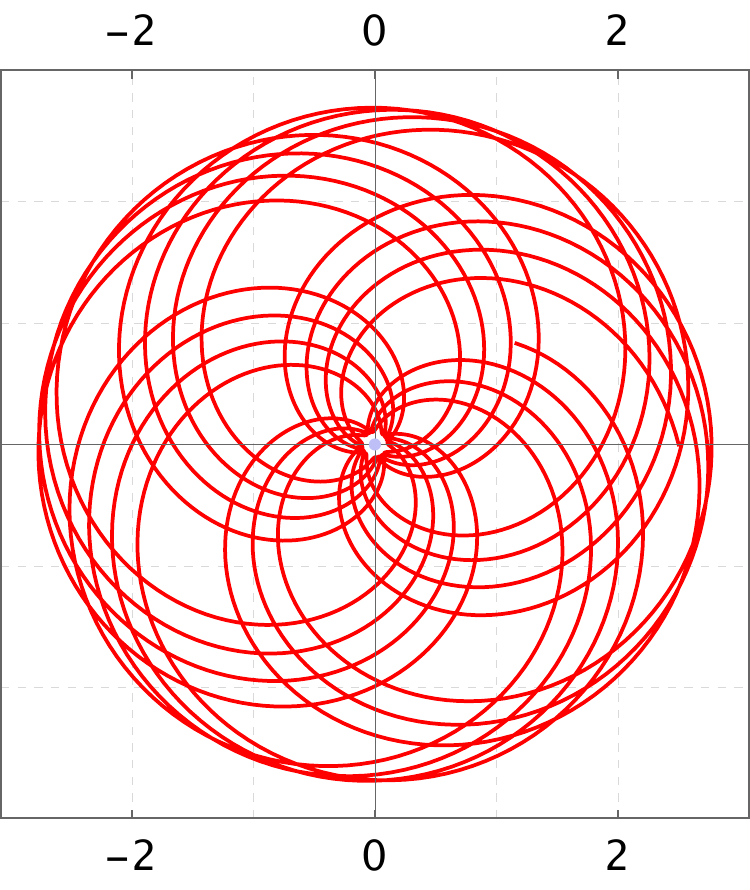}\quad\quad
    \includegraphics[width=0.25\linewidth]{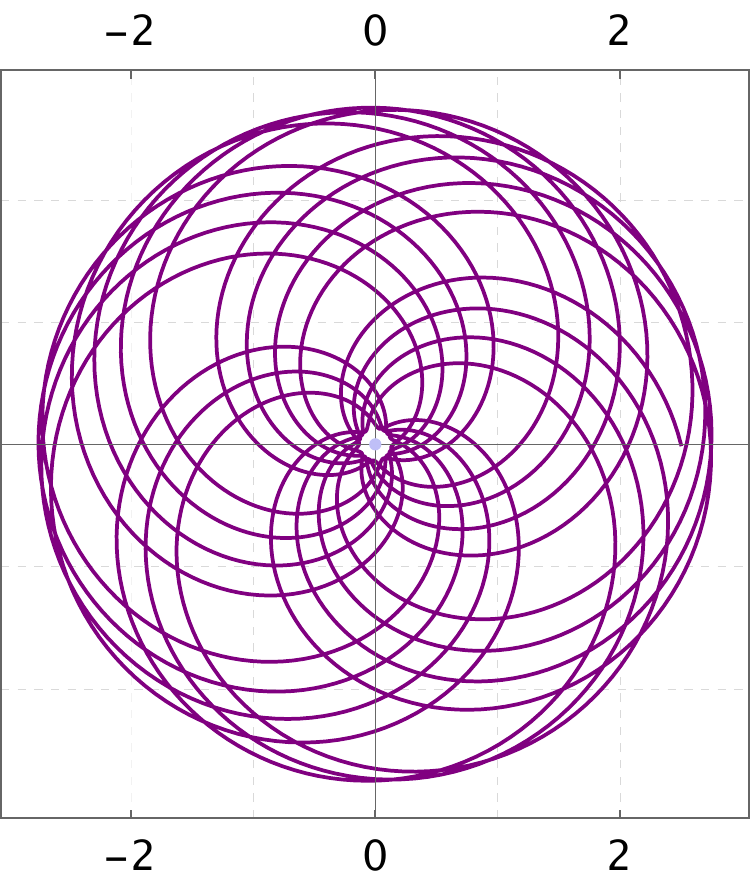}\quad\quad
    \includegraphics[width=0.25\linewidth]{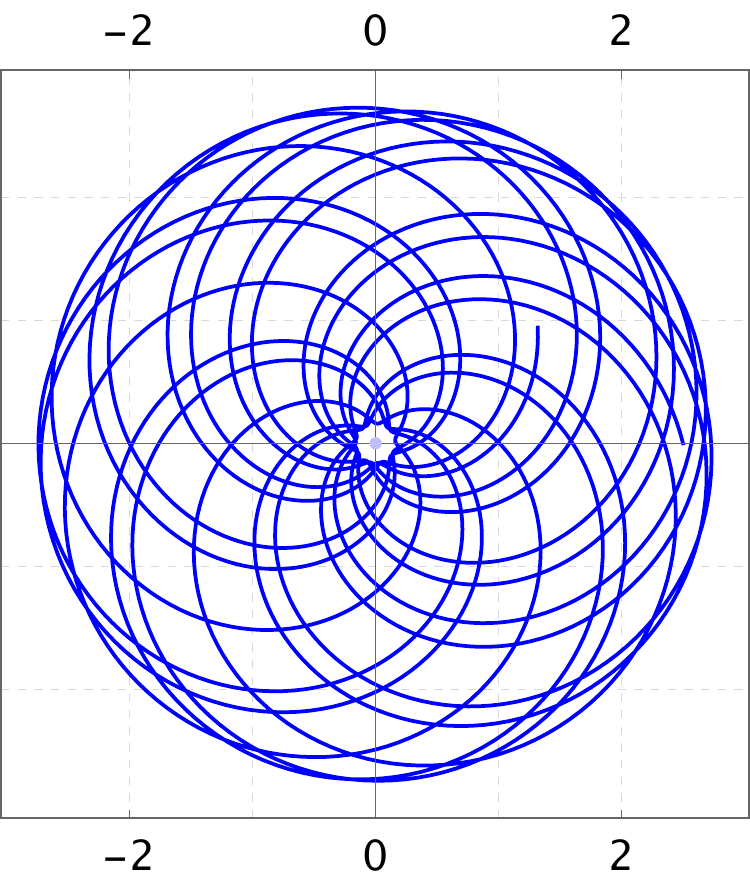}\\
    (g) $\alpha=0.12, \lambda=0.20$ \hspace{2cm} (h) $\alpha=0.15, \lambda=0.25$ \hspace{2cm} (i) $\alpha=0.18, \lambda=0.30$
    \caption{\footnotesize Illustration of the photon trajectories (\ref{bb9}) for different values the string parameter \(\alpha\), NC parameter \(\lambda\), and their combination. The first row corresponds to $\lambda=0.2, \mathrm{N}=0.1$, the second row to $\alpha=0.1, \mathrm{N}=0.1$, and the third row to $\mathrm{N}=0.1$. Throughout these panels, we set $M=1,\,w=-2/3$.}
    \label{fig:trajectory}
\end{figure}

Equation(\ref{bb8}) represents the photon trajectory equation in the chosen charged AdS BH solution coupled with a cloud of strings surrounded by  quintessence-like fluid in noncommutative geometry background. We observed that the photon trajectories depends on various geometrical parameters. These include  string clouds characterized by the parameter \( \alpha \), quintessential field parameters \((\mathrm{N},w)\), and non-commutative geometry effects characterized by the parameter \( \lambda \).

For a specific state parameter, for example, $w=-2/3$, the photon trajectories from Eq. (\ref{bb8}) reduces as
\begin{eqnarray}
    \frac{d^2u}{d\phi^2}+(1-\alpha)\,u =3\,M\,u^2-2\,\lambda\,M\,u^3+\frac{1}{2}\,\mathrm{N}.\label{bb99}  
\end{eqnarray}

Figure \ref{fig:trajectory} illustration the photon trajectories (\ref{bb9}) for different values the string parameter \(\alpha\), NC parameter \(\lambda\), and their combination. 

\begin{figure}[ht!]
    \centering
    \includegraphics[width=0.3\linewidth]{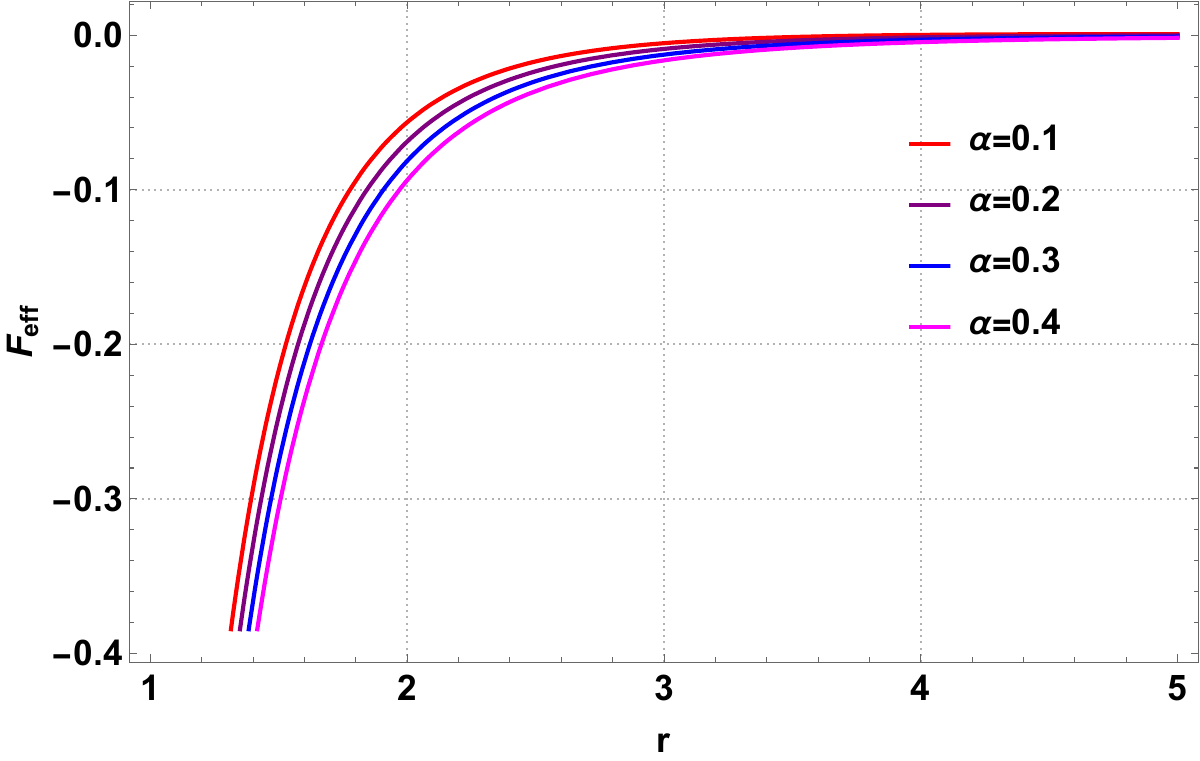}\quad\quad
    \includegraphics[width=0.3\linewidth]{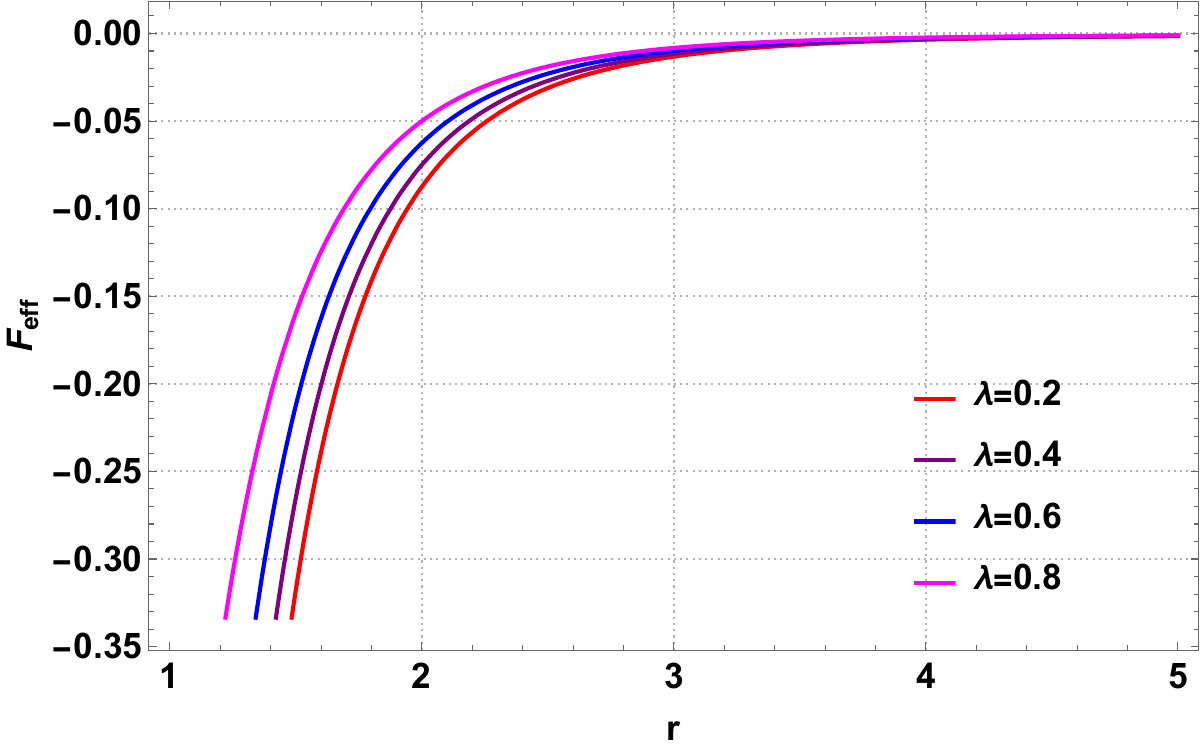}\quad\quad
    \includegraphics[width=0.3\linewidth]{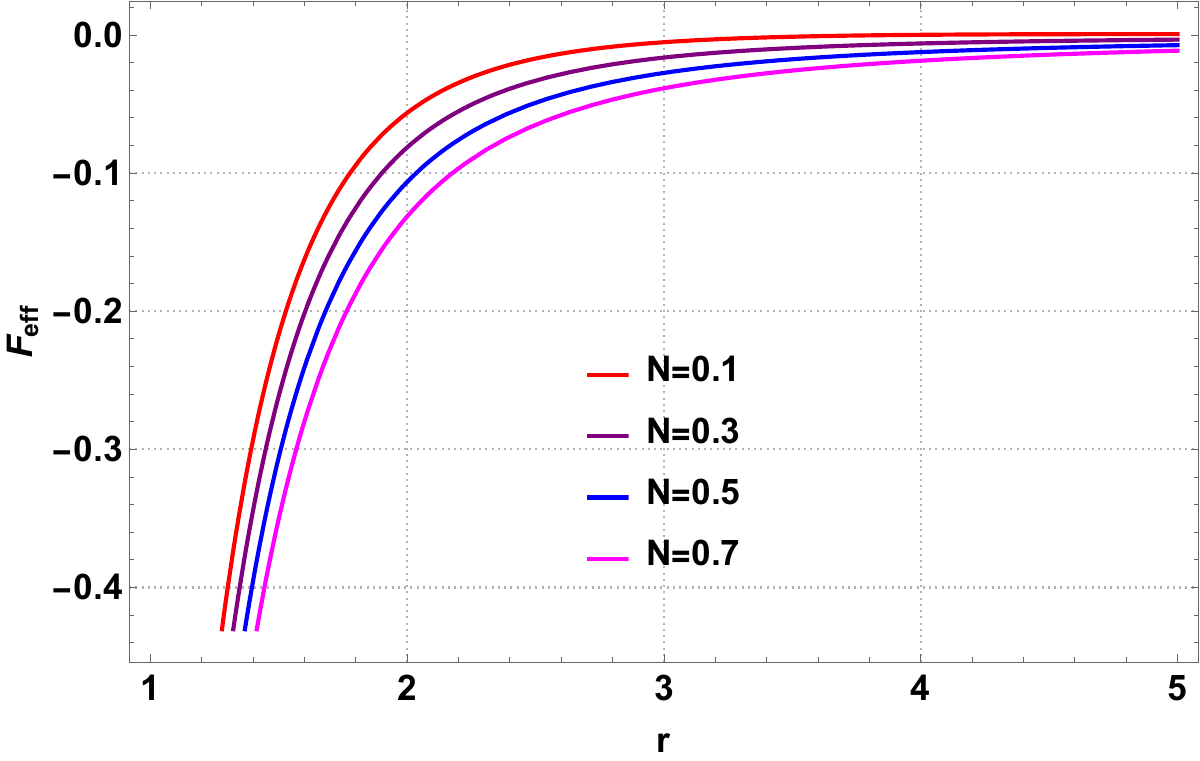}\\
    (a) $\lambda=0.5,\,\mathrm{N}=0.1$ \hspace{4cm} (b) $\alpha=0.1,\,\mathrm{N}=0.2$ \hspace{4cm} (c) $\lambda=0.5,\,\alpha=0.1$
    \caption{\footnotesize Behavior of the effective force (\ref{bb999}) experienced by the photon particles for different values of the string parameter \(\alpha\), NC parameter \(\lambda\), and the normalization constant $\mathrm{N}$. Here, we set $M=1=\mathrm{L}, w=-2/3$.}
    \label{fig:force}
\end{figure}

\begin{center}
    {\bf Effective radial force on photon particles}
\end{center}

We now compute the force acting on the massless photon particles as they traverse the gravitational field of the considered BH. This analysis highlights the influence of the additional parameters the cloud of strings parameter \( \alpha \) and the noncommutative parameter \( a \) on the gravitational force, in comparison to the standard Schwarzschild case. The force on a photon is derived from the effective potential \( V_{\text{eff}} \), which governs radial motion in curved spacetime. For null geodesics, the force is given by
\begin{eqnarray}
    \mathrm{F}_\text{ph}(r)=-\frac{1}{2}\frac{dV_\text{eff}(r)}{dr}=\frac{\mathrm{L}^2}{r^3}\,\left[1 - \alpha - \frac{3\,M}{r} + \frac{2\,\lambda\, M}{r^2} - \frac{(3\,w + 3)\,\mathrm{N}}{2\,r^{3\,w+1}}\right].\label{bb9}
\end{eqnarray}

From the above expression (\ref{bb9}), it is clear that the force acting on the photon particles under the influence of the gravitational field depends on various geometrical parameters. These include  string clouds characterized by the parameter \( \alpha \), quintessential field characterized by the parameters \(\mathrm{N},w\), the noncommutative geometry effects characterized by the parameter \( a \), the BH mass \( M \), and the angular momentum \( \mathrm{L} \).

For a specific state parameter, for example, $w=-2/3$, the effective radial force Eq. (\ref{bb9}) reduces as
\begin{eqnarray}
    \mathrm{F}_\text{ph}(r)=\frac{\mathrm{L}^2}{r^3}\,\left[1 - \alpha - \frac{3\,M}{r} + \frac{2\,\lambda\, M}{r^2} - \frac{\mathrm{N}}{2}\,r\right].\label{bb999}
\end{eqnarray}

In Figure \ref{fig:force}, we illustrate the effective radial force experienced by the photon particles as a function of $r$ for different values of the string parameter \(\alpha\), NC geometry parameter \(\lambda\), and the normalization constant \(\mathrm{N}\) associated with the quintessence field. Panels (a) and (c) reveal that the effective force decreases as \(\alpha\) and \(\mathrm{N}\) increase, respectively. Conversely, panel (b) demonstrates that the effective force increases with increasing values of the NC geometry parameter \(\lambda\), suggesting a stronger gravitational influence in this regime.

\begin{center}
    {\bf Photon sphere radius and BH Shadow}
\end{center}

For circular photon orbits having radius $r = r_\text{ph}$, the effective potential satisfies the following conditions
\begin{equation}
    \mathrm{E}^2=V_\text{eff}(r),\quad\quad     V'_\text{eff}(r)=0,\quad\quad    V''_\text{eff}(r)<0,\label{cc1}
\end{equation}
where prime denotes partial derivative w. r. t. argument.

Using the first condition, we find the critical impact parameter for photons as,
\begin{eqnarray}
   \frac{1}{\beta_c}=\frac{\sqrt{1-\alpha - \frac{2\,M}{r_\text{ph}}+\frac{\lambda\,M}{r_\text{ph}^2}-\frac{\mathrm{N}}{r_\text{ph}^{3\,w+1}} - \frac{\Lambda}{3}\,r_\text{ph}^2}}{r_\text{ph}}.\label{cc2}
\end{eqnarray}\label{cc22}
When $\beta < \beta_c$, photon particles are captured by the black hole. At the critical value $\beta = \beta_c$, photons follow unstable circular orbits at the photon sphere. For $\beta > \beta_c$, photon particles are able to escape to infinity.

Using the second condition, one can determine the radius of photon sphere satisfying the following relation
\begin{eqnarray}
    2\,\mathcal{F}=r\,\mathcal{F}'\Rightarrow 1 - \alpha - \frac{3\,M}{r} + \frac{2\,\lambda\, M}{r^2} - \frac{(3\,w + 3)\,\mathrm{N}}{2\,r^{3\,w+1}}=0.\label{cc3}
\end{eqnarray}
Without setting state parameter $w$, it is very difficult to determine radius of the photon sphere. For an example, setting $w=-2/3$, from relation (\ref{cc3}), we find
\begin{eqnarray}
    2\,(1 - \alpha)\,r^2 - 6\,M\,r + 4\,\lambda\,M-\mathrm{N}\,r^3=0,\label{cc4}
\end{eqnarray}
a cubic equation in $r$ whose real valued solution gives us the radius of the photon sphere. However, by setting suitable values of $\lambda$, $\alpha$ and $\mathrm{N}$, one can determine numerical results for this photon sphere radius. Table \ref{tab:1} presented one such examples of numerical values for $\alpha=0.1, 0.2, 0.3$.

\begin{table}[ht!]
\centering
\small
\begin{minipage}{0.3\textwidth}
\centering
\begin{tabular}{|c|c|c|c|}
\hline
\(\alpha\) & \(\lambda\) & \(N\) & \(r_\text{ph}\) \\
\hline
\multirow{12}{*}{0.1} 
 & 0.2 & 0.2 & 0.139044 \\
 & 0.2 & 0.3 & 0.138995 \\
 & 0.2 & 0.4 & 0.138946 \\
 & 0.4 & 0.2 & 0.291299 \\
 & 0.4 & 0.3 & 0.290808 \\
 & 0.4 & 0.4 & 0.290321 \\
 & 0.6 & 0.2 & 0.460316 \\
 & 0.6 & 0.3 & 0.458166 \\
 & 0.6 & 0.4 & 0.456078 \\
 & 0.8 & 0.2 & 0.651425 \\
 & 0.8 & 0.3 & 0.644591 \\
 & 0.8 & 0.4 & 0.638191 \\
\hline
\end{tabular}
\caption*{(a) \(\alpha=0.1\)}
\end{minipage}%
\hfill
\begin{minipage}{0.3\textwidth}
\centering
\begin{tabular}{|c|c|c|c|}
\hline
\(\alpha\) & \(\lambda\) & \(N\) & \(r_\text{ph}\) \\
\hline
\multirow{12}{*}{0.2}
 & 0.2 & 0.2 & 0.138349 \\
 & 0.2 & 0.3 & 0.138302 \\
 & 0.2 & 0.4 & 0.138254 \\
 & 0.4 & 0.2 & 0.287987 \\
 & 0.4 & 0.3 & 0.287523 \\
 & 0.4 & 0.4 & 0.287065 \\
 & 0.6 & 0.2 & 0.451234 \\
 & 0.6 & 0.3 & 0.449296 \\
 & 0.6 & 0.4 & 0.447409 \\
 & 0.8 & 0.2 & 0.631192 \\
 & 0.8 & 0.3 & 0.625404 \\
 & 0.8 & 0.4 & 0.619935 \\
\hline
\end{tabular}
\caption*{(b) \(\alpha=0.2\)}
\end{minipage}%
\hfill
\begin{minipage}{0.3\textwidth}
\centering
\begin{tabular}{|c|c|c|c|}
\hline
\(\alpha\) & \(\lambda\) & \(N\) & \(r_\text{ph}\) \\
\hline
\multirow{12}{*}{0.3}
 & 0.2 & 0.2 & 0.137669 \\
 & 0.2 & 0.3 & 0.137622 \\
 & 0.2 & 0.4 & 0.137576 \\
 & 0.4 & 0.2 & 0.284826 \\
 & 0.4 & 0.3 & 0.284388 \\
 & 0.4 & 0.4 & 0.283954 \\
 & 0.6 & 0.2 & 0.442869 \\
 & 0.6 & 0.3 & 0.44111\\
 & 0.6 & 0.4 & 0.439393 \\
 & 0.8 & 0.2 & 0.613445 \\
 & 0.8 & 0.3 & 0.608454 \\
 & 0.8 & 0.4 & 0.603706 \\
\hline
\end{tabular}
\caption*{ (c) \(\alpha=0.3\)}
\end{minipage}
\caption{\footnotesize Numerical real positive roots of the photon sphere radius \(r_\text{ph}\) given in Eq.~ (\ref{cc4}) for different values of \(\lambda\) and $\mathrm{N}$. Here $M=1$.}
\label{tab:1}
\end{table}

Next, we determine the BH shadow radius and is given by the following relation:
\begin{equation}
    R_\text{ps}=\beta_c=\frac{r_\text{ph}}{\sqrt{1-\alpha - \frac{2\,M}{r_\text{ph}}+\frac{\lambda\,M}{r_\text{ph}^2}-\frac{\mathrm{N}}{r_\text{ph}^{3\,w+1}} - \frac{\Lambda}{3}\,r_\text{ph}^2}}.\label{cc5}
\end{equation}
For an example, setting the state parameter $w=-2/3$, we find the shadow radius as
\begin{equation}
    R_\text{ps}=\frac{r_\text{ph}}{\sqrt{1-\alpha - \frac{2\,M}{r_\text{ph}}+\frac{\lambda\,M}{r_\text{ph}^2}-\mathrm{N}\,r_\text{ph}- \frac{\Lambda}{3}\,r_\text{ph}^2}}.\label{cc6}
\end{equation}
From the above expression (\ref{cc6}), it becomes evident that the shadow radius depends on various geometrical parameters. These include  the string clouds characterized by the parameter \( \alpha \), the normalization constant \(\mathrm{N}\), the noncommutative geometry effects characterized by the parameter \( a \).

\subsection{Time-like Geodesics}

In this subsection, we analyze the time-like geodesics in the given BH space-time geometry. Our focus is to derive the equations governing the motion of massive test particles, which provide crucial insights into the orbital dynamics influenced by the gravitational field of the BH. A key aspect of this analysis is the determination of the innermost stable circular orbit (ISCO) radius, which represents the smallest radius at which a massive particle can maintain a stable circular orbit around the BH. The ISCO plays a fundamental role in astrophysical phenomena such as accretion disk dynamics, black hole shadow imaging, and gravitational wave emission from inspiraling compact objects.

Our study explores how the BH parameters-including mass, string clouds, NV geometry and the cosmological constant affect the ISCO radius. Understanding these effects enhances our comprehension of particle dynamics in strong gravitational fields and provides observational signatures to distinguish between different BH models.

\begin{figure}[ht!]
    \centering
    \includegraphics[width=0.3\linewidth]{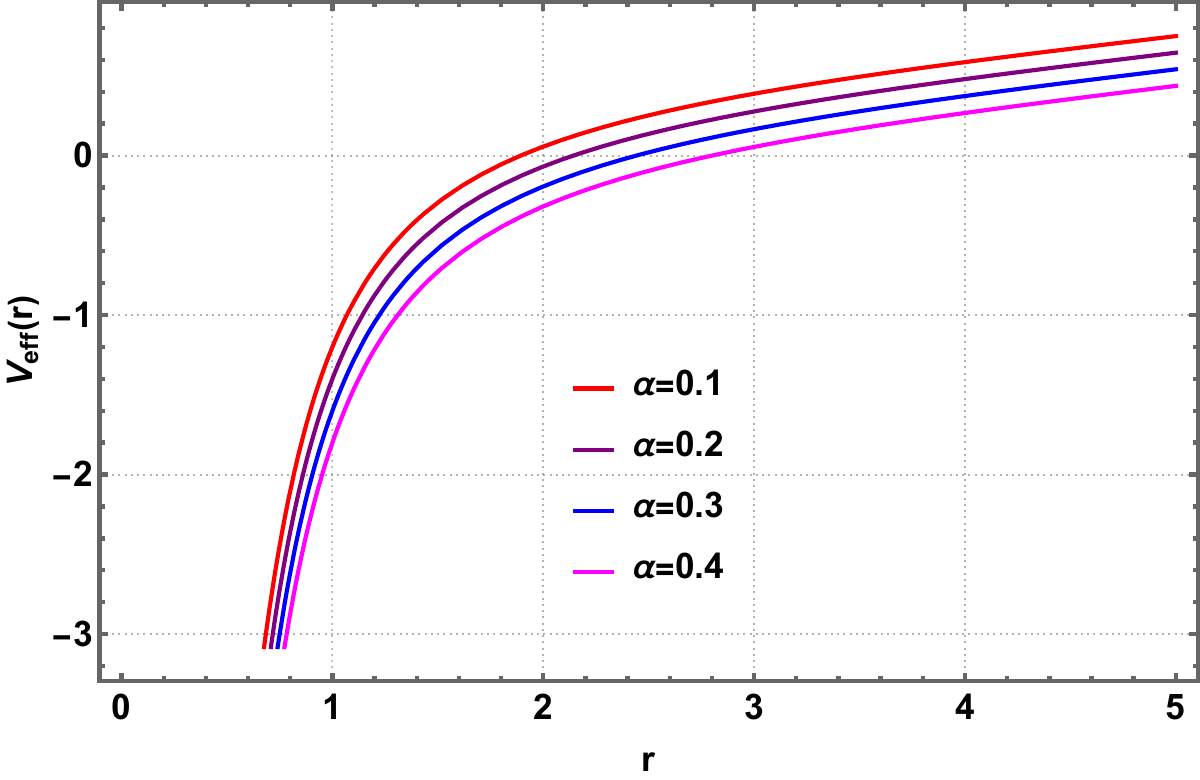}\quad\quad
    \includegraphics[width=0.3\linewidth]{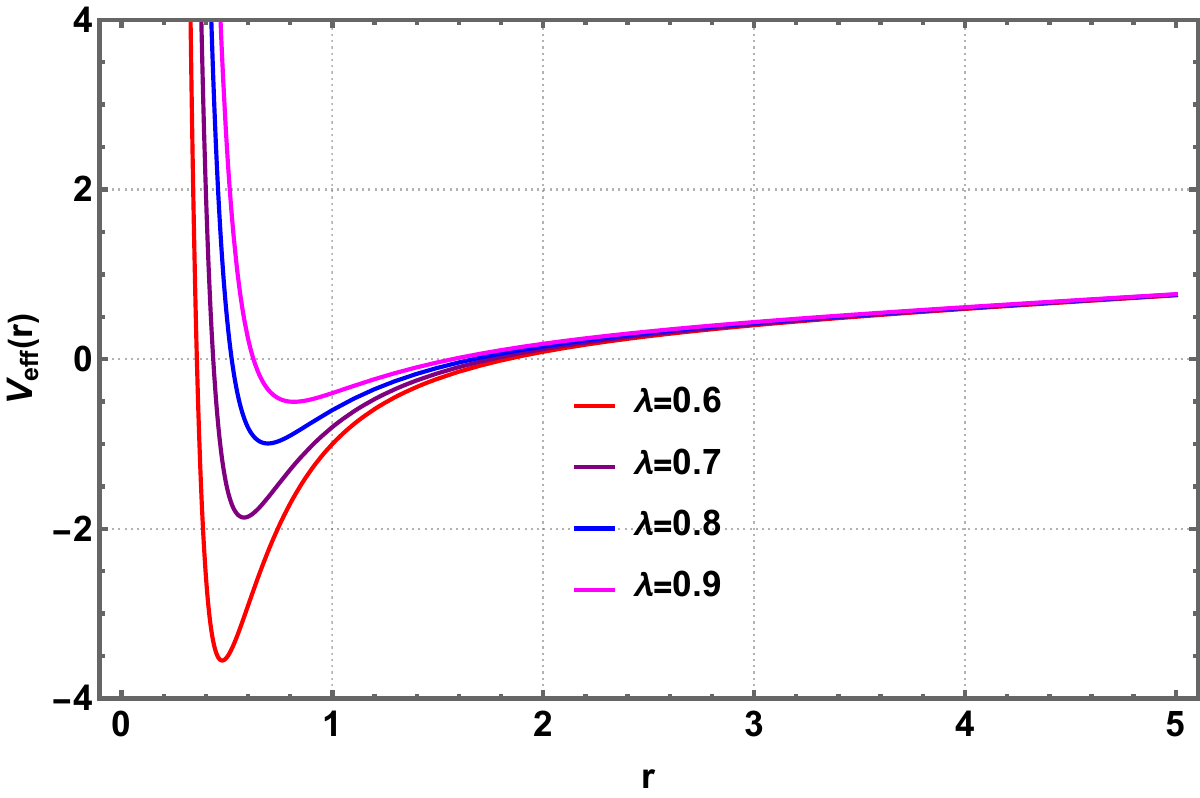}\quad\quad
    \includegraphics[width=0.3\linewidth]{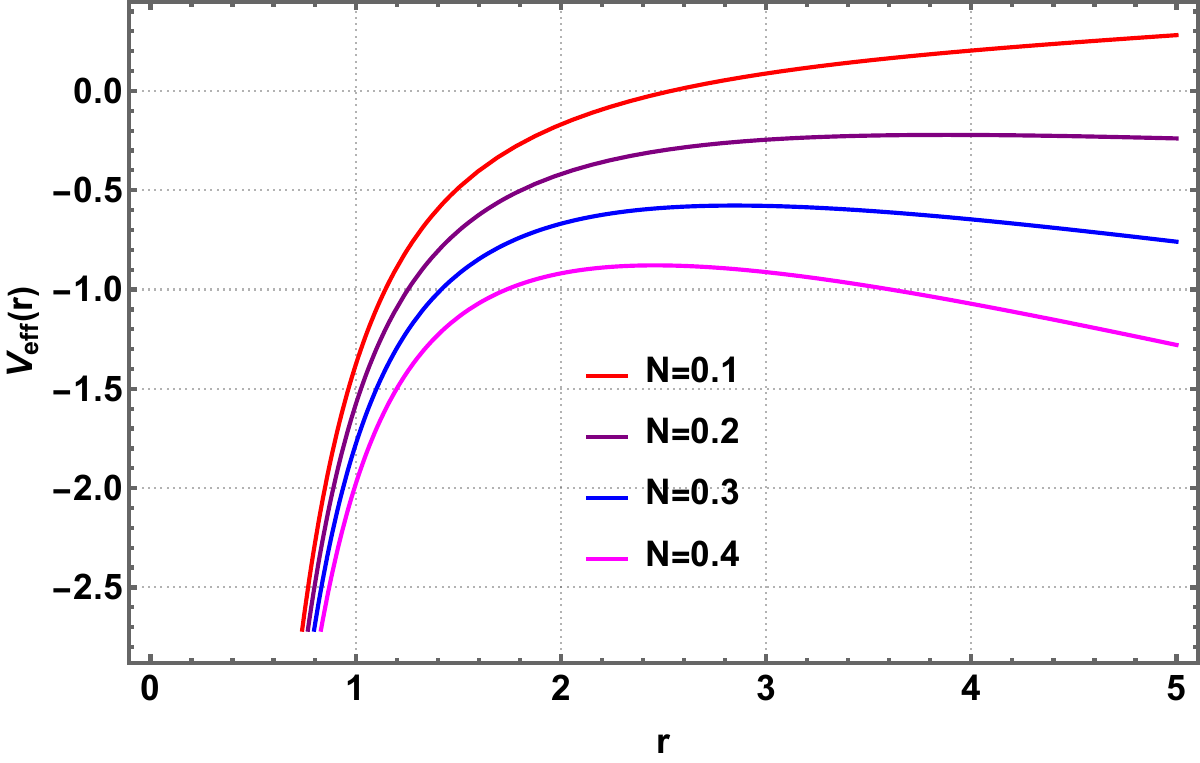}\\
    (a) $\lambda=0.5,\,\mathrm{N}=0.01$ \hspace{4cm} (b) $\alpha=0.1,\,\mathrm{N}=0.01$ \hspace{4cm} (c) $\lambda=0.5,\,\alpha=0.1$
    \caption{\footnotesize Behavior of the effective potential for the null geodesic by varying values of the string parameter \(\alpha\), NC parameter \(\lambda\), and the normalization constant $\mathrm{N}$. Here, we set $M=1=\mathrm{L}, w=-2/3$.}
    \label{fig:time-like}
\end{figure}

In Figure \ref{fig:time-like}, we present the effective potential for time-like geodesic motion as functions of $r$ for different values of the string parameter \(\alpha\), NC geometry parameter \(\lambda\), and the normalization constant \(\mathrm{N}\) associated with the quintessence field. Panels (a) and (c) show  that the effective potential decreases as \(\alpha\) and \(\mathrm{N}\) increase, respectively, indicating that higher values of these parameters lower the potential barrier. Conversely, panel (b) demonstrates that the effective potential increases with increasing values of the NC geometry parameter \(\lambda\), suggesting a stronger gravitational influence by this parameter.

For time-like geodesics, we have $\epsilon=-1$, and hence, the effective potential Eq. (\ref{bb5}) for time-like geodesic reduces as,
\begin{eqnarray}
     V_\text{eff}(r)=\mathcal{F}\,\left(1+\frac{\mathrm{L}^2}{\,r^2}\right)=\left(\epsilon+\frac{\mathrm{L}^2}{\,r^2}\right)\,\left[1-\alpha - \frac{2\,M}{r}+\frac{\lambda\,M}{r^2}-\frac{\mathrm{N}}{r^{3\,w+1}} - \frac{\Lambda}{3}\,r^2\right]. \label{dd1}
\end{eqnarray}

For innermost stable circular orbits, following conditions must be satisfied:
\begin{equation}
    \mathrm{E}^2=V_\text{eff}(r),\quad\quad V'_\text{eff}(r)=0,\quad\quad V''_\text{eff}(r) \geq 0,\label{dd2}
\end{equation}
prime denotes ordinary derivative w. r. t. $r$.

Using the above effective potential, the find relation in terms of the metric function as,
\begin{equation}
    2\,\mathcal{F}(r)\,\mathcal{F}''(r) + 6\, \mathcal{F}(r)\,\mathcal{F}')(r)/r - 4\, (\mathcal{F}'(r))^2 =0.\label{dd5}
\end{equation}
Substituting the metric function, we find the following polynomial equation: {\small
\begin{align}
&\left(1 - \alpha - \frac{2M}{r} + \frac{\lambda M}{r^2} - \frac{N}{r^{3w+1}} - \frac{\Lambda}{3}r^2\right)
\left(-\frac{4M}{r^3} + \frac{6\lambda M}{r^4} - \frac{(3w+1)(3w+2)N}{r^{3w+3}} - \frac{2\Lambda}{3} \right)+ \frac{3}{r} \left(1 - \alpha - \frac{2M}{r} + \frac{\lambda M}{r^2} - \frac{N}{r^{3w+1}} - \frac{\Lambda}{3}r^2\right)\times\nonumber\\
&\left(\frac{2M}{r^2} - \frac{2\lambda M}{r^3} + \frac{(3w+1)N}{r^{3w+2}} - \frac{2\Lambda}{3}r \right)- 2\left(\frac{2M}{r^2} - \frac{2\lambda M}{r^3} + \frac{(3w+1)N}{r^{3w+2}} - \frac{2\Lambda}{3}r\right)^2 = 0.\label{dd6}
\end{align}
}

\begin{table}[ht!]
\centering
\small
\begin{tabular}{|c|c|c|c|}
\hline
\text{CS parameter} ($\alpha$) & \text{Constant} ($\mathrm{N}$) & \text{NC parameter} ($\lambda$) & \text{larger root } \(r=r_{\text{ISCO}} > 0\) \\
\hline
0.1 & 0.2 & 0.2 & 6.6068 \\
\hline
0.1 & 0.2 & 0.3 & 6.5729 \\
\hline
0.1 & 0.3 & 0.2 & 9.4963 \\
\hline
0.1 & 0.3 & 0.3 & 9.4769 \\
\hline
0.2 & 0.2 & 0.2 & 6.7246 \\
\hline
0.2 & 0.2 & 0.3 & 6.7039 \\
\hline
0.2 & 0.3 & 0.2 & 9.7055 \\
\hline
0.2 & 0.3 & 0.3 & 9.6909 \\
\hline
0.3 & 0.2 & 0.2 & 6.7868 \\
\hline
0.3 & 0.2 & 0.3 & 6.7719 \\
\hline
0.3 & 0.3 & 0.2 & 9.8605 \\
\hline
0.3 & 0.3 & 0.3 & 9.8489 \\
\hline
\end{tabular}
\caption{\footnotesize Numerical results for the larger positive root \(r\) for various CS parameter \(\alpha\), normalization constant \(\mathrm{N}\), and NC geometry parameter \(\lambda\). Here, we set $M=1, \Lambda=-0.03$.}
\label{tab:2}
\end{table}

For a state parameter $w=-2/3$, we find the following polynomial equation for ISCO radius as,
\begin{align}
&\left(1 - \alpha - \frac{2M}{r} + \frac{\lambda M}{r^2} - N r - \frac{\Lambda}{3} r^2 \right) 
\left(-\frac{4M}{r^3} + \frac{6 \lambda M}{r^4} - \frac{2 \Lambda}{3} \right) \nonumber\\
&+ \frac{3}{r} \left(1 - \alpha - \frac{2M}{r} + \frac{\lambda M}{r^2} - N r - \frac{\Lambda}{3} r^2 \right)
\left(\frac{2M}{r^2} - \frac{2 \lambda M}{r^3} - N - \frac{2 \Lambda}{3} r \right)- 2\left(\frac{2M}{r^2} - \frac{2 \lambda M}{r^3} - N - \frac{2 \Lambda}{3} r \right)^2 = 0.\label{dd7}
\end{align}
Solving Eq.~(\ref{dd7}) analytically is a challenging task and therefore, we provided numerical values by selecting values of various geometrical parameters involved in the space-time geometry (see Table \ref{tab:2}).

For circular orbits, employing the conditions $\mathrm{E}^2=V_\text{eff}(r)$ and $V'_\text{eff}(r)=0$, we find the specific angular momentum and energy of timelike particles as,
\begin{align}
\mathrm{L}_\text{spc.}=\frac{
2 M r - 2 \lambda M + (3w + 1) N r^{1 - 3w} - \frac{2 \Lambda}{3} r^4
}{
2(1-\alpha) - \frac{6 M}{r} + \frac{4 \lambda M}{r^2} - \frac{(3w + 3) N}{r^{3w + 1}}
},\quad\quad 
\mathrm{E}_\text{spc.}=\frac{\left(1-\alpha - \frac{2\,M}{r}+\frac{\lambda\,M}{r^2}-\frac{\mathrm{N}}{r^{3\,w+1}} - \frac{\Lambda}{3}\,r^2\right)^2}{\sqrt{2(1-\alpha) - \frac{6 M}{r} + \frac{4 \lambda M}{r^2} - \frac{(3w + 3) N}{r^{3w + 1}}}}.\label{dd8}
\end{align}
For a specific state parameter, $w=-2/3$, we find
\begin{align}
\mathrm{L}_\text{spc.}=\frac{
2 M r - 2 \lambda M -N r^3- \frac{2 \Lambda}{3} r^4
}{
2(1-\alpha) - \frac{6 M}{r} + \frac{4 \lambda M}{r^2} -N\,r},\quad\quad
\mathrm{E}_\text{spc.}&=\frac{\left(1-\alpha - \frac{2\,M}{r}+\frac{\lambda\,M}{r^2}-N\,r - \frac{\Lambda}{3}\,r^2\right)^2}{\sqrt{2(1-\alpha) - \frac{6 M}{r} + \frac{4 \lambda M}{r^2} -N\,r}}.\label{dd9}
\end{align}

From the above expression (\ref{dd8}), it becomes clear that the specific angular momentum and energy of massive test particles in circular orbits under the influence of the gravitational field depends on various geometrical parameters. These include  string clouds characterized by the parameter \( \alpha \), quintessential field characterized by the parameters \(\mathrm{N}, w\), the non-commutative geometry effects characterized by the parameter \( a \), and the BH mass \( M \)).

\section{Thermodynamic Properties of Black Hole}\label{sec:3}

In this section, we investigate the thermodynamic properties of the Schwarzschild-AdS BH modified by a Lorentzian noncommutative geometry and surrounded by a cloud of strings. We derive key quantities such as the Hawking temperature, entropy, specific heat, and Gibbs free energy to analyze the BH’s stability and phase structure. Additionally, we examine how variations in the non-commutative and strings cloud parameters influence thermodynamic behavior and critical phenomena, highlighting the interplay between geometric deformation and string contributions in a generalized BH setting.

The metric of the selected Schwarzschild-AdS BH in noncommutative geometry with a cloud of strings is given by
\begin{equation}
     \mathcal{F} (r)=1-\alpha - \frac{2\,M}{r}+\frac{\lambda\,M}{r^2}-\frac{\mathrm{N}}{r^{3\,w+1}} - \frac{\Lambda}{3}\,r^2.\label{ff1}
\end{equation}

To study the thermodynamic properties of the selected BH, we first determine the BH ADS mass in terms of event horizon $r_{+}$. This is given by
\begin{eqnarray}
   1-\alpha - \frac{2\,M}{r_{+}}+\frac{\lambda\,M}{r_{+}^2}-\frac{\mathrm{N}}{r_{+}^{3\,w+1}} - \frac{\Lambda}{3}\,r_{+}^2=0,         
\end{eqnarray}
which brings us to
\begin{eqnarray}
    M = \Bigg(\frac{r_{+}^2}{\lambda-2r_{+}}\Bigg)\Bigg(\alpha-1  +\frac{\mathrm{N}}{r_{+}^{3\,w+1}} + \frac{\Lambda}{3}\,r_{+}^2\Bigg).          \label{ff3}
\end{eqnarray}

The Hawking temperature of a BH is defined by its surface gravity as follows \cite{FA4,FA6,FA7}:
\begin{eqnarray}
   T_H=\frac{\kappa}{2\,\pi}=\frac{\mathcal{F}'(r_{+})}{4\,\pi}=\frac{3N(1+3\omega)r_{+}-(3M\lambda-3M \,r_{+}+\Lambda r_{+}^4)\,2r_{+}^{3\omega}}{12\pi\, r_{+}^{3(\omega+1)}}\label{ff5} 
\end{eqnarray}

For a static spherically symmetric black hole, the entropy can be calculated by the area law, $S=A/4=\pi\, r_{+}^2$, which is a monotonic function of the horizon radius. The specific heat capacity is defined by
\begin{eqnarray}
    C_p=T_H\,\frac{\delta\,S}{\delta\,T_H}=2\,\pi\,r_{+}\,\frac{T_H}{\frac{\delta\,T_H}{\delta\,r_{+}}}\label{ff9}
\end{eqnarray}
Substituting the Hawking temperature from Eq. (\ref{ff5}) in the Eq. (\ref{ff9}), we find  
\begin{eqnarray}
    C_p=\Bigg[\frac{(3M\lambda-3M \,r_{+}+\Lambda r_{+}^4)\,2r_{+}^{3\omega}-3N(1+3\omega)r_{+}}{3N(2+9\omega+9\omega^2)r_{+}+(6M'r_{+}-9M'\lambda+\Lambda r_{+}^4)2r_{+}^{3\omega}}\Bigg]2\pi\,r_{+}^2,
\end{eqnarray}
where $M'\equiv{\frac{\delta\,M}{\delta\,r_{+}}}$.

Now, we focus on the Gibb's free energy of the thermodynamic system. The Gibbs free energy is defined by
\begin{eqnarray}
   G=M-T_{H}S=M-\frac{3N(1+3\omega)r_{+}^3-(3M\lambda-3M \,r_{+}+\Lambda r_{+}^4)\,2r_{+}^{3\omega+2}}{6\, r_{+}^{3(\omega+1)}}.\label{ff11}
\end{eqnarray}

\begin{figure}[ht!]
\centering
\includegraphics[height=5cm,width=7cm]{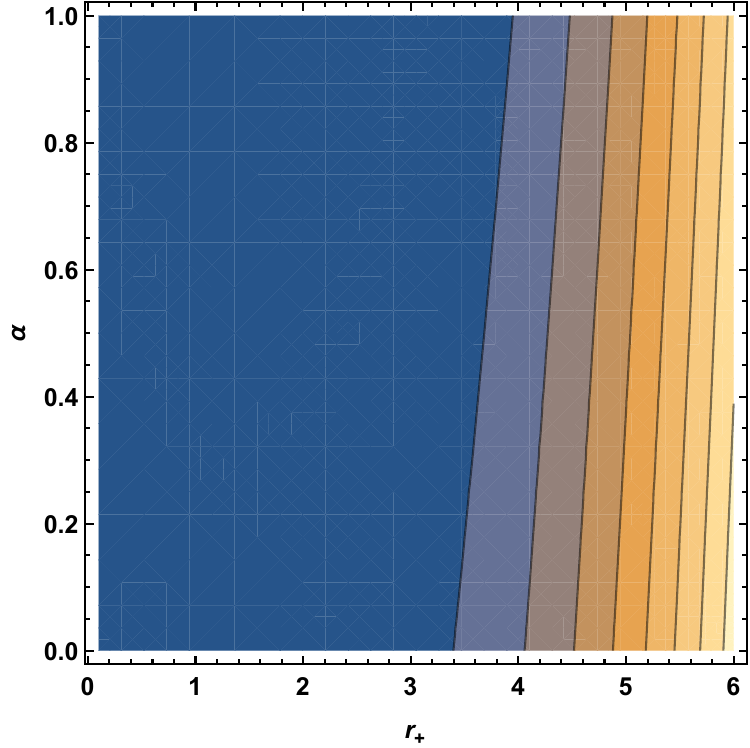}
\includegraphics[height=5cm,width=7cm]{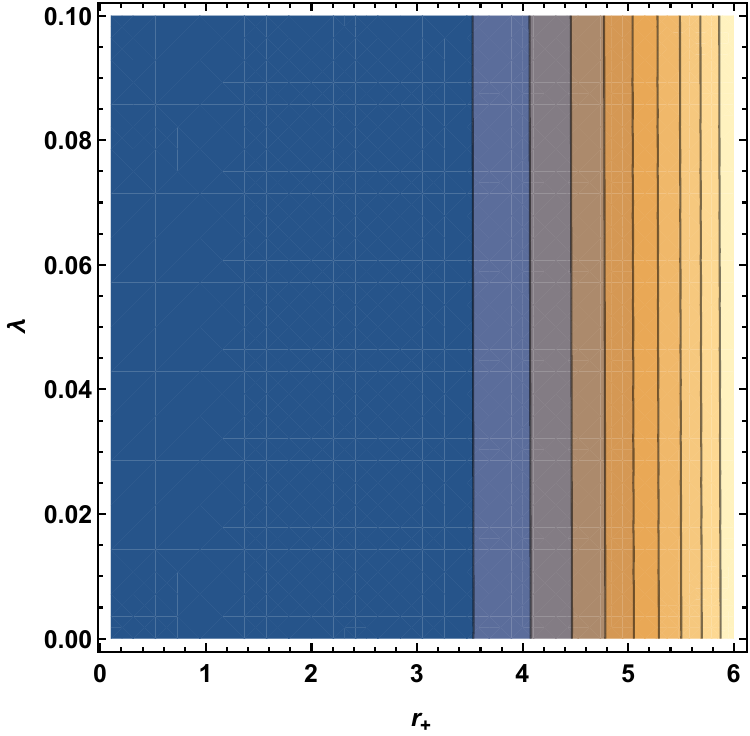}\\
(a) \hspace{6cm} (b)\\
\includegraphics[height=5cm,width=7cm]{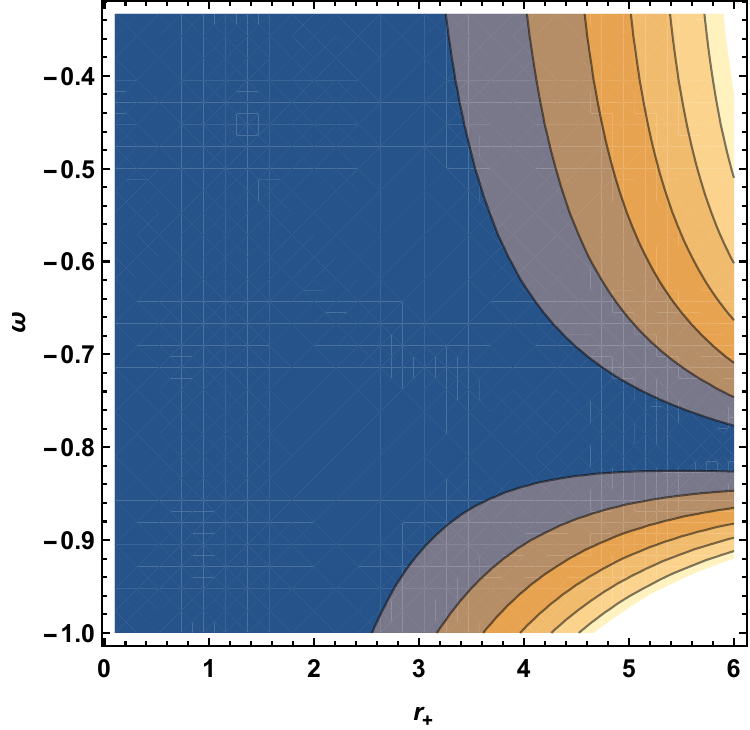}
\includegraphics[height=5cm,width=7cm]{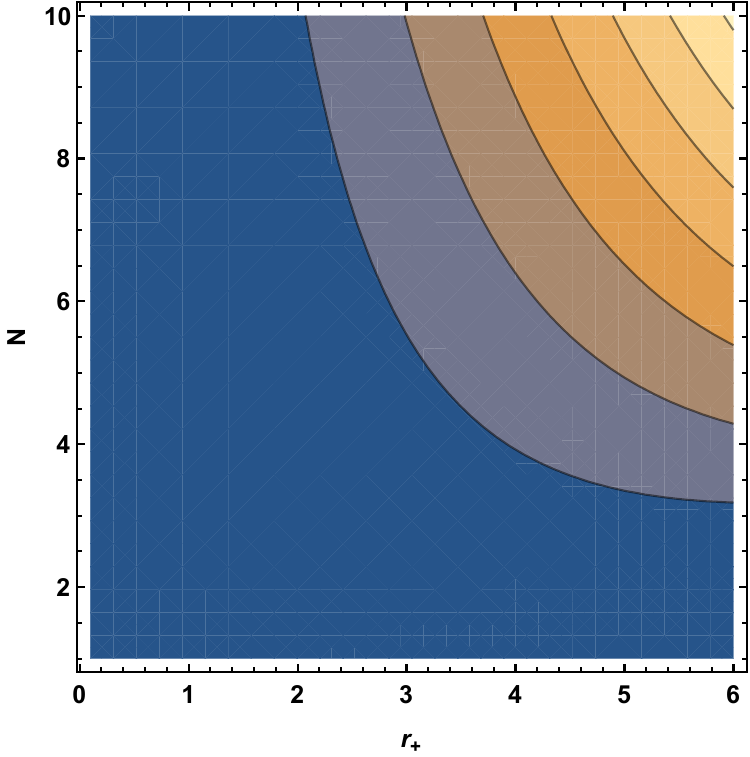}\\
(c) \hspace{6cm} (d)
\caption{ADS mass $M(r_+)$. (a) $\Lambda=-1$, $\lambda=0.1$, $\omega=-2/3$ and $N=1$. (b) $\Lambda=-1$, $\alpha=0.3$, $\omega=-2/3$ and $N=1$. (c) $\Lambda=-1$, $\alpha=0.3$, $\lambda=0.1$ and $\omega=-2/3$. (d) $\Lambda=-1$, $\alpha=0.3$, $\lambda=0.1$ and $N=1$.}
\label{fig7}
\end{figure}

Figure~\ref{fig7} shows contour plots of the AdS black hole mass $M(r_+)$ as a function of the event horizon radius $r_+$ and various model parameters. In Fig.\ref{fig7} (a), we observe that increasing the string cloud parameter $\alpha$ results in a monotonic increase in the BH mass, particularly for large $r_+$. This suggests that the string cloud contributes an effective positive energy density that enhances the total mass. In Fig.\ref{fig7} (b), the noncommutative geometry parameter $\lambda$ is varied, and the results reveal that for small horizon radii, the BH mass is nearly insensitive to changes in $\lambda$, but for larger $r_+$, increasing $\lambda$ significantly raises $M$. This reflects the growing influence of noncommutative corrections in the IR regime. In Fig.\ref{fig7} (c), we examine the effect of the quintessence equation-of-state parameter $\omega$, and find that as $\omega$ approaches $-1$, the mass decreases for fixed $r_+$, in agreement with the interpretation of quintessence as a repulsive dark energy component. The contour structure indicates a nontrivial interplay between $r_+$ and $\omega$, especially in the small-radius regime, where mass minima appear. Finally, Fig.\ref{fig7} (d) varies the normalization parameter $N$, showing that larger values of $N$ lead to a rapid increase in mass for small $r_+$, due to the stronger contribution from the quintessence field. These results underscore how the combined effects of $\alpha$, $\lambda$, $N$, and $\omega$ govern the effective energy content of the space-time and thereby modulate both gravitational strength and thermodynamic behavior.

\begin{figure}[ht!]
\centering
\includegraphics[height=5cm,width=7cm]{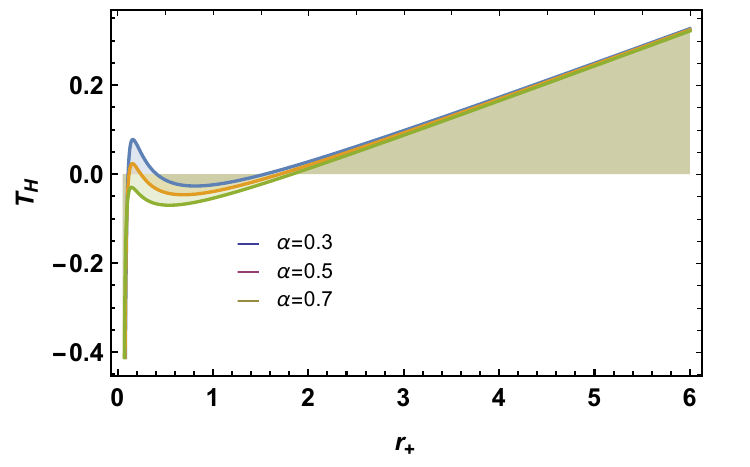}
\includegraphics[height=5cm,width=7cm]{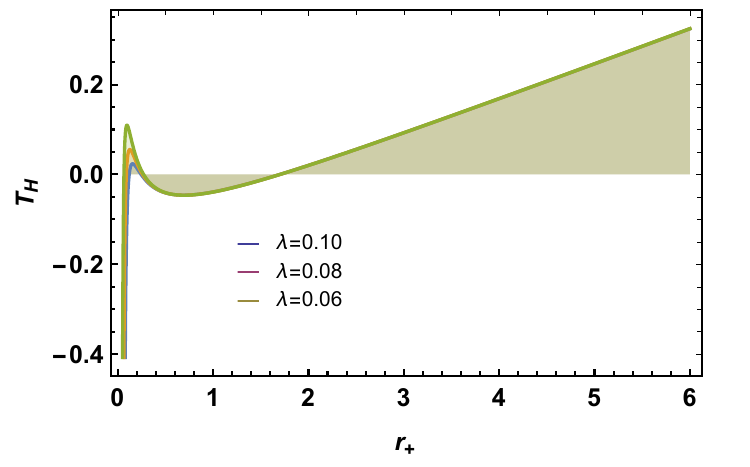}\\
(a) \hspace{6cm} (b)\\
\includegraphics[height=5cm,width=7cm]{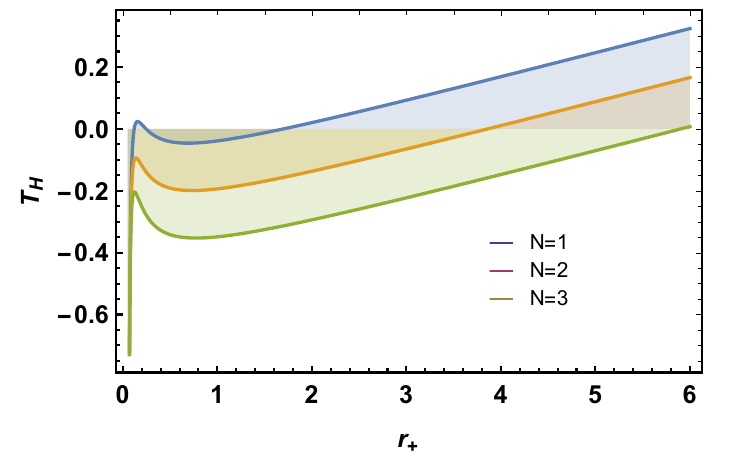}
\includegraphics[height=5cm,width=7cm]{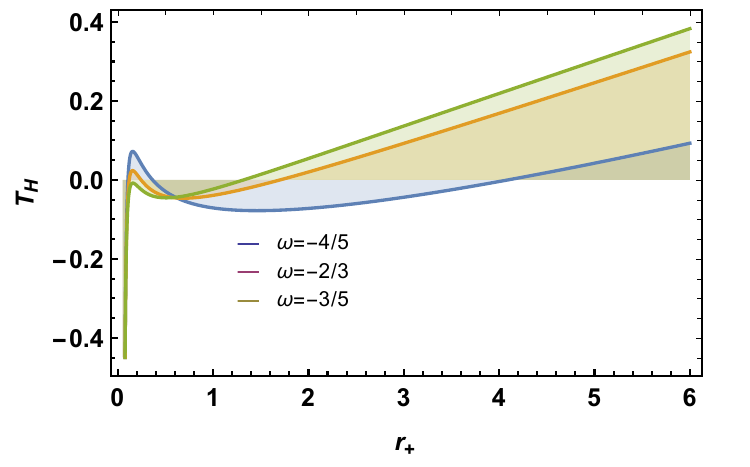}\\
(c) \hspace{6cm} (d)
\caption{Hawking Temperature with $M=1$. (a) $\Lambda=-1$, $\lambda=0.1$, $\omega=-2/3$ and $N=1$. (b) $\Lambda=-1$, $\alpha=0.3$, $\omega=-2/3$ and $N=1$. (c) $\Lambda=-1$, $\alpha=0.3$, $\lambda=0.1$ and $\omega=-2/3$. (d) $\Lambda=-1$, $\alpha=0.3$, $\lambda=0.1$ and $N=1$.}
\label{fig1}
\end{figure}

Figure~\ref{fig1} illustrates the behavior of the Hawking temperature $T_H$ as a function of the event horizon radius $r_+$ for different values of the model parameters. In Fig.\ref{fig1} (a), we observe that increasing the string cloud parameter $\alpha$ causes the temperature to rise more steeply for large $r_+$ while also shifting the position of the minimum temperature toward smaller horizon radii. This reflects the impact of the string tension on the surface gravity and the thermodynamic stability. Fig.\ref{fig1} (b) explores the influence of the noncommutative parameter $\lambda$, where higher values of $\lambda$ slightly suppress the temperature at small $r_+$, confirming the regularizing role of noncommutative geometry in the ultraviolet regime. In Fig.\ref{fig1} (c), increasing the quintessence normalization parameter $N$ systematically lowers the Hawking temperature across all values of $r_+$, particularly reducing the local minimum and delaying the onset of the positive-temperature region, which may signal delayed thermodynamic stability. Fig.\ref{fig1} (d) reveals the effect of varying the equation-of-state parameter $\omega$, showing that more negative values (closer to $\omega = -1$) decrease the temperature in the small-horizon regime and smooth out the temperature gradient. Across all cases, we note the presence of a local minimum in $T_H(r_+)$, a typical signature of phase transition points. These results collectively demonstrate that the temperature profile—and consequently the thermodynamic stability—is highly sensitive to the interplay between noncommutative effects, string cloud tension, and quintessence parameters, emphasizing the need for a full multi-parameter analysis of BH thermodynamics in such deformed AdS backgrounds.

\begin{figure}[ht!]
\centering
\includegraphics[height=5cm,width=7cm]{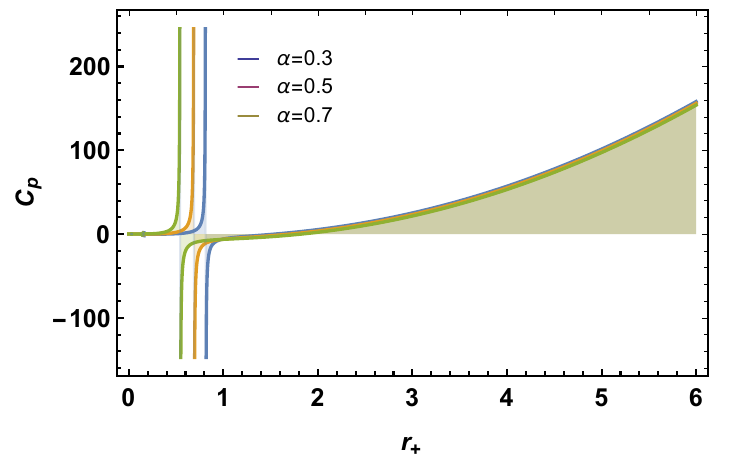}
\includegraphics[height=5cm,width=7cm]{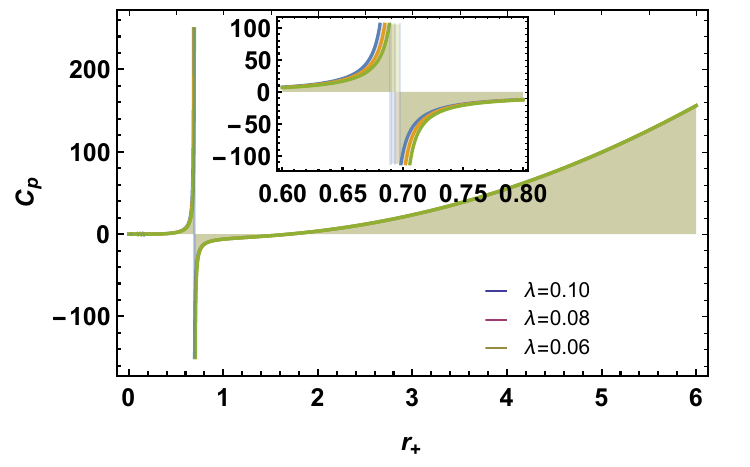}\\
(a) \hspace{6cm} (b)\\
\includegraphics[height=5cm,width=7cm]{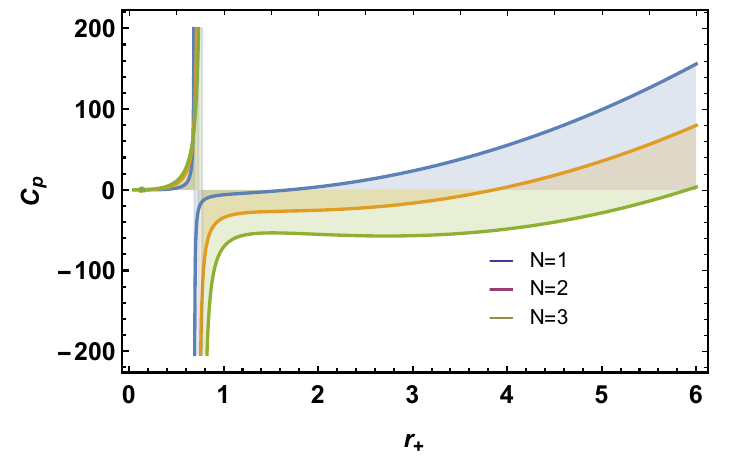}
\includegraphics[height=5cm,width=7cm]{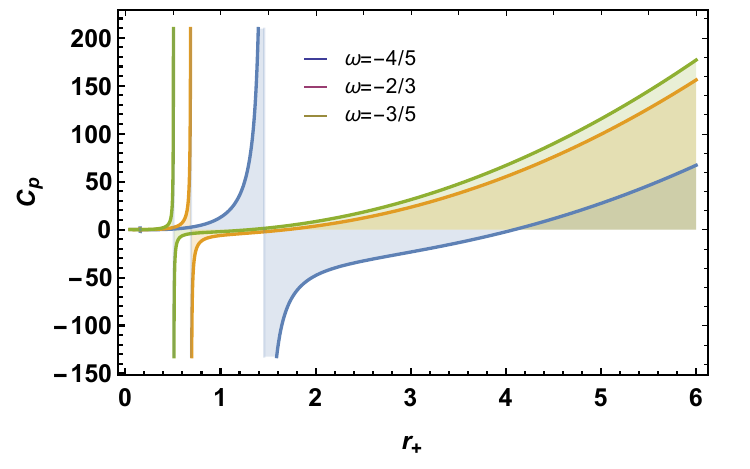}\\
(c) \hspace{6cm} (d)
\caption{Heat capacity. (a) $\Lambda=-1$, $\lambda=0.1$, $\omega=-2/3$ and $N=1$. (b) $\Lambda=-1$, $\alpha=0.3$, $\omega=-2/3$ and $N=1$. (c) $\Lambda=-1$, $\alpha=0.3$, $\lambda=0.1$ and $\omega=-2/3$. (d) $\Lambda=-1$, $\alpha=0.3$, $\lambda=0.1$ and $N=1$.}
\label{fig2}
\end{figure}

Figure~\ref{fig2} presents the behavior of the heat capacity $C_p$ as a function of the event horizon radius $r_+$ for various model parameters, providing insight into the thermodynamic stability of the black hole. In Fig.\ref{fig2} (a), increasing the string cloud parameter $\alpha$ causes a shift in the divergence point of $C_p$ toward smaller values of $r_+$, indicating an earlier onset of the phase transition. The presence of vertical asymptotes signals second-order phase transitions, which are characteristic of thermodynamically unstable to stable transitions. Fig.\ref{fig2} (b) reveals that the noncommutative parameter $\lambda$ influences the location and sharpness of the divergence. Higher values of $\lambda$ slightly delay the divergence and smooth the transition, consistent with its regularizing role. In Fig.\ref{fig2} (c), increasing the normalization parameter $N$ of the quintessence field pushes the divergence to larger radii and elevates the post-transition values of $C_p$, showing that strong quintessence contributions enhance thermal stability in the large BH phase. Fig.\ref{fig2} (d) explores the dependence on the equation-of-state parameter $\omega$; more negative values of $\omega$ lead to an earlier divergence and a steeper gradient of $C_p$, amplifying thermal instability in the small-$r_+$ regime. Overall, these results confirm that the system undergoes a typical small-to-large black hole phase transition. The position and nature of the heat capacity divergence are highly sensitive to the combined effects of $\alpha$, $\lambda$, $N$, and $\omega$, emphasizing their role in determining the black hole's phase structure and thermodynamic stability.

\begin{figure}[ht!]
\centering
\includegraphics[height=5cm,width=7cm]{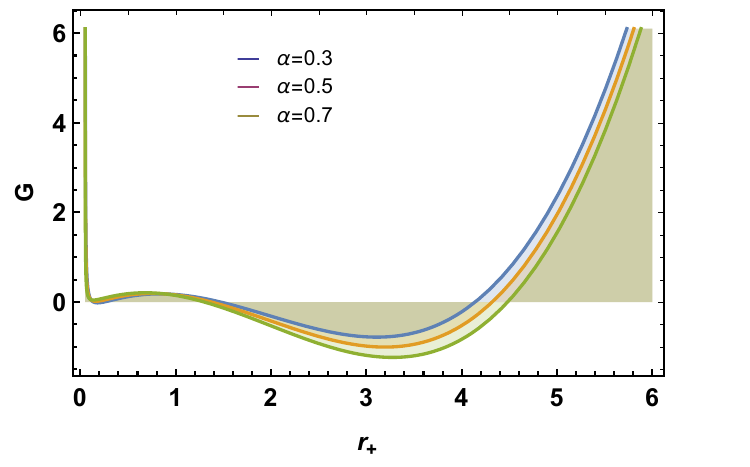}
\includegraphics[height=5cm,width=7cm]{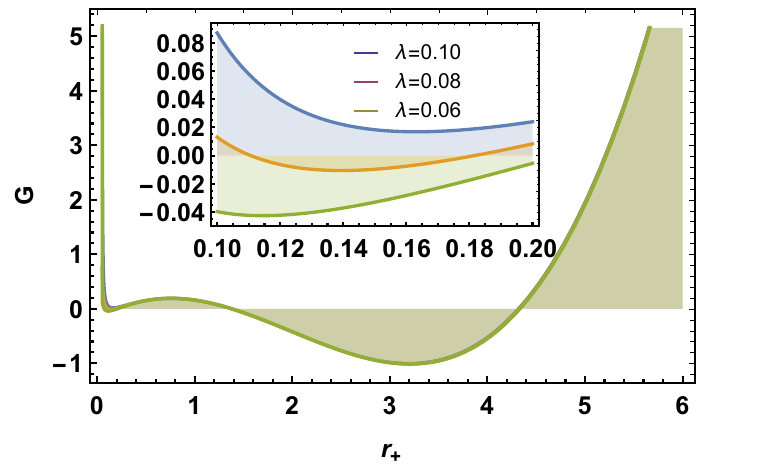}\\
(a) \hspace{6cm} (b)\\
\includegraphics[height=5cm,width=7cm]{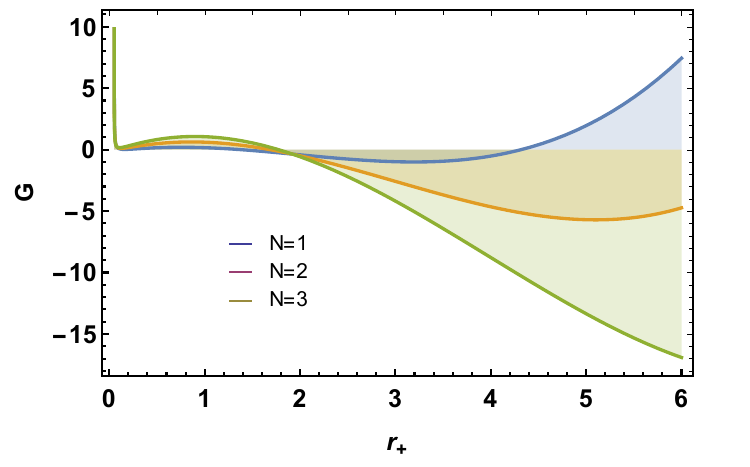}
\includegraphics[height=5cm,width=7cm]{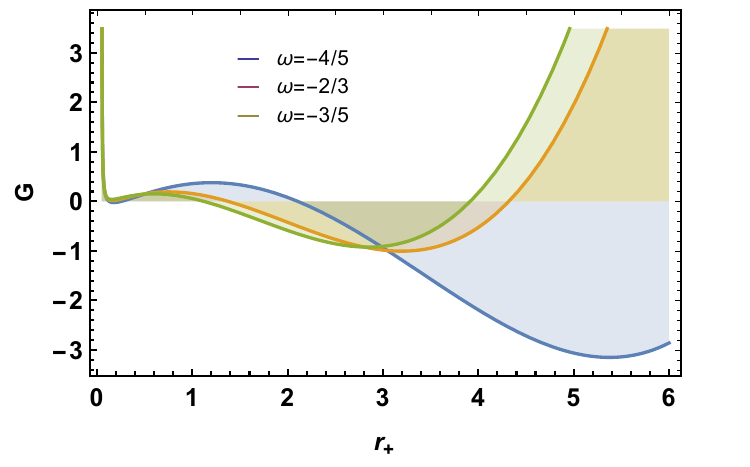}\\
(c) \hspace{6cm} (d)
\caption{Gibb's free energy. (a) $\Lambda=-1$, $\lambda=0.1$, $\omega=-2/3$ and $N=1$. (b) $\Lambda=-1$, $\alpha=0.3$, $\omega=-2/3$ and $N=1$. (c) $\Lambda=-1$, $\alpha=0.3$, $\lambda=0.1$ and $\omega=-2/3$. (d) $\Lambda=-1$, $\alpha=0.3$, $\lambda=0.1$ and $N=1$.}
\label{fig6}
\end{figure}

Figure~\ref{fig6} displays the behavior of the Gibbs free energy $G$ as a function of the event horizon radius $r_+$ for different values of the model parameters. This thermodynamic potential plays a crucial role in identifying global stability and phase transitions in black hole systems. In Fig.\ref{fig6} (a), increasing the string cloud parameter $\alpha$ raises the Gibbs free energy across all $r_+$ and shifts the minimum of $G$ toward smaller horizon radii. A well-defined global minimum emerges, indicating a possible Hawking-Page-type phase transition.  Fig.\ref{fig6} (b) shows that the noncommutative parameter $\lambda$ has a significant impact only in the small-$r_+$ regime, where higher values of $\lambda$ lift the Gibbs energy, smoothing the transition near the origin. This is consistent with the ultraviolet regularization induced by noncommutativity. In  Fig.\ref{fig6} (c), the normalization parameter $N$ of the quintessence field is varied; larger values of $N$ lower the Gibbs free energy minimum and shift it toward larger radii, reflecting the repulsive effect of the dark energy component. Finally,  Fig.\ref{fig6} (d) demonstrates the role of the equation-of-state parameter $\omega$. As $\omega$ becomes more negative, the minimum of $G$ becomes deeper and occurs at smaller $r_+$, further enhancing the likelihood of a small-to-large black hole transition. In all cases, the presence of a swallowtail-like structure or non-monotonic behavior in $G(r_+)$ indicates the possibility of first-order phase transitions. These results reinforce that the interplay between $\alpha$, $\lambda$, $N$, and $\omega$ governs not only the local but also the global thermodynamic stability of the noncommutative AdS black hole system.

\section{Scalar Perturbations of Black Hole}\label{sec:4}

In this part, perturbations of zero-spin scalar field by deriving the Klein-Gordon equation around the selected BH solution is studied. The dynamics of massless scalar field, denoted as $\Phi$ are governed by the covariant form of the Klein-Gordon equation , which describes the evolution of a scalar field in a curved spacetime. This equation serves as the fundamental equation of motion for the scalar field in a gravitational background. Numerous authors have been studied the scalar perturbation in various BH solutions in the literature (see, for examples \cite{FA1,FA2,FA3,FA4,FA5,FA6,FA7,FA8,FA9,FA10,FA11,FA12}). 

The Klein-Gordon equation for a massless scalar field in a general curved spacetime is given by the following wave equation
\begin{equation}
    \frac{1}{\sqrt g}\,\partial_\mu(\sqrt-g\,g^{\mu\nu}\partial_\nu\Phi)=0,\label{hh1}
\end{equation}
Where $g_{\mu\nu}$ is the metric tensor, g = det$(g_{\mu\nu})$ is the determinant of the metric tensor and $\delta_\mu$ is the partial derivative with respect to the coordinate system. 

Moreover, we consider an ansatz for the scalar field $\Phi(t,r,\theta,\phi)$ of the following form:
\begin{equation}
    \Phi(t,r,\theta,\phi)=\exp(-i\omega\,t)\,Y^m_{\ell}(\theta,\phi)\,\frac{\psi(r)}{r},\label{hh2}
 \end{equation}
Where $Y^m_{l}(\theta,\phi)$ are the spherical harmonics, $\omega$ is the (possibly complex) temporal frequency in the Fourier domain and $\psi(r)$ is a propagating scalar field in the curved spacetime.

Explicitly writing Eq. (\ref{hh1}) using Eq. (\ref{aa3}) and Eq. (\ref{hh2}), we find the Schrodinger-like wave equation:
\begin{equation}
    \frac{d^2\psi(r_*)}{dr_*^2}+(\omega^2-\mathcal{V})\psi(r_*)=0,\label{hh3}
\end{equation}
where the tortoise coordinate $r_{*}$ is defined by
\begin{equation}
    r_*=\int\,\frac{dr}{\mathcal F(r)}.\label{hh4}
\end{equation}
The scalar perturbative potential is given by the following expression
\begin{align}
    \mathcal{V}(r)=\left(1-\alpha - \frac{2\,M}{r}+\frac{\lambda\,M}{r^2}-\frac{\mathrm{N}}{r^{3\,w+1}} - \frac{\Lambda}{3}\,r^2\right)\,\left(\frac{\ell\,(\ell+1)}{r^2}+\frac{2M}{r^3} - \frac{2 \lambda M}{r^4} + N (3w + 1) r^{-3w - 3} - \frac{2 \Lambda}{3}\right).\label{hh5}
\end{align}

For a specific state parameter, $w=-2/3$, the perturbative potential reduces as,
\begin{equation}
    \mathcal{V}(r) = \left( 1 - \alpha - \frac{2M}{r} + \frac{\lambda M}{r^2} - N r - \frac{\Lambda}{3} r^2 \right)
\left( \frac{\ell(\ell+1)}{r^2} + \frac{2M}{r^3} - \frac{2 \lambda M}{r^4} - \frac{N}{r} - \frac{2 \Lambda}{3} \right).\label{hh6}
\end{equation}

\begin{figure}[ht!]
    \centering
    \includegraphics[width=0.3\linewidth]{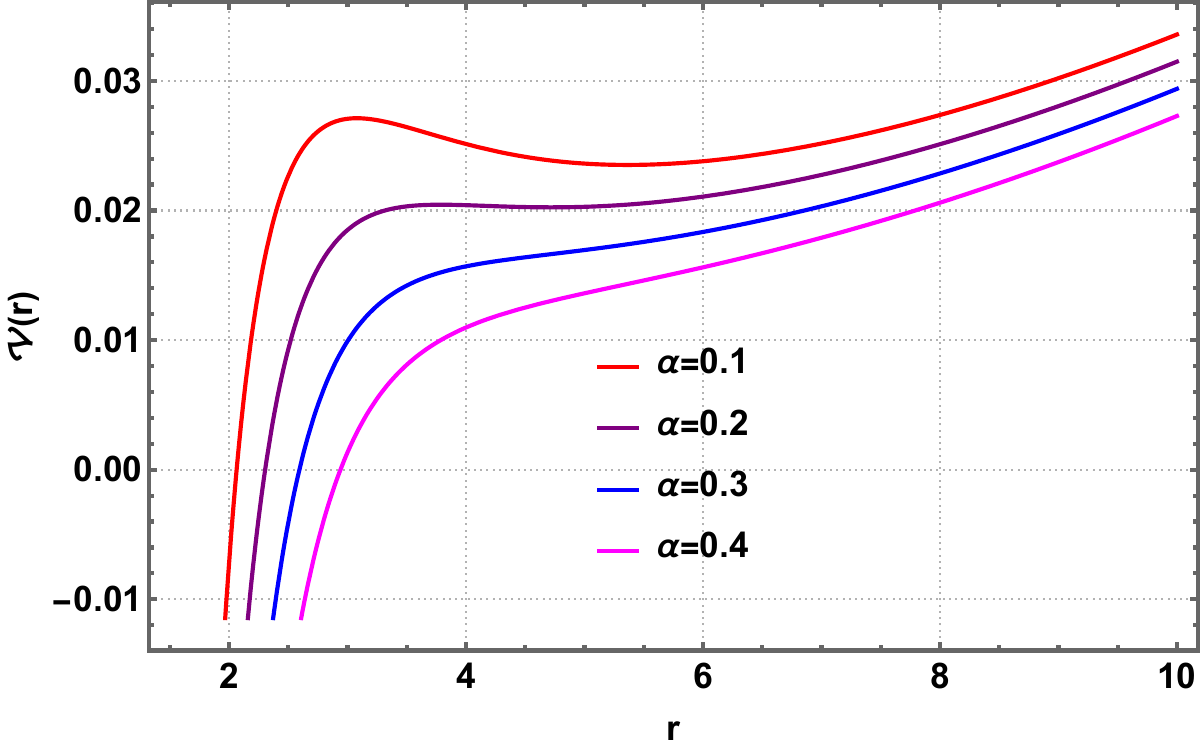}\quad\quad
    \includegraphics[width=0.3\linewidth]{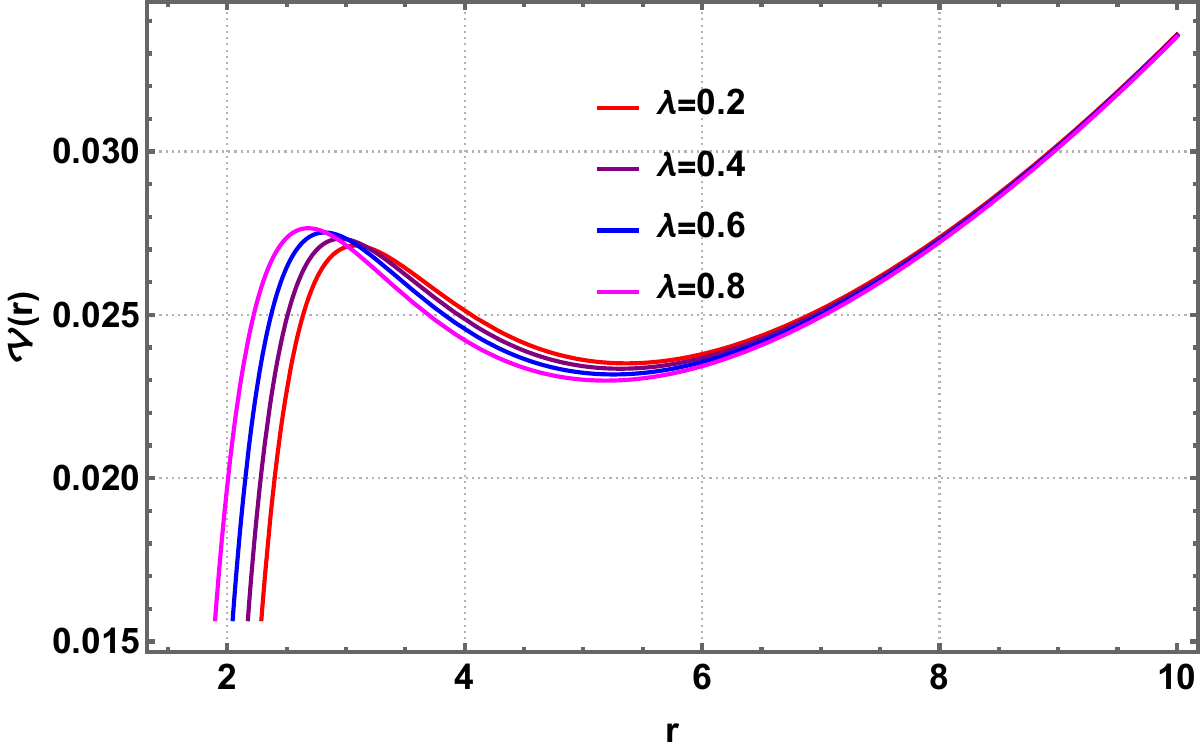}\quad\quad
    \includegraphics[width=0.3\linewidth]{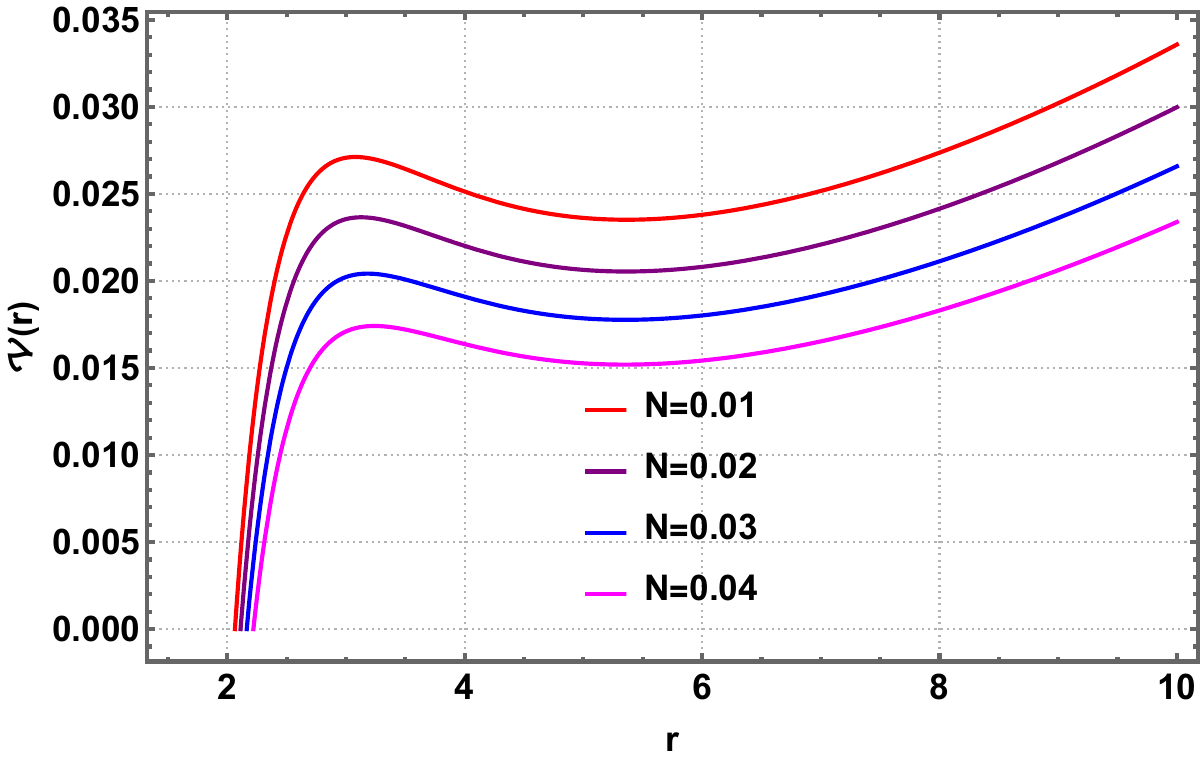}\\
    (a) $\lambda=0.2,\,\mathrm{N}=0.01$ \hspace{4cm} (b) $\alpha=0.1,\,\mathrm{N}=0.01$\hspace{4cm} (c) $\alpha=0.1,\,\lambda=0.2$
    \caption{\footnotesize Behavior of the scalar perturbative potential given in Eq.~(\ref{hh6}) as a function of $r$ by varying values of the string parameter \(\alpha\), NC parameter \(\lambda\), and the normalization constant $\mathrm{N}$. Here, we set $M=1, w=-2/3, \Lambda=-0.03$.}
    \label{fig:12}
\end{figure}

From the above expression (\ref{hh5}), it becomes clear that the perturbative scalar potential is influenced by various geometrical parameters. These include string clouds characterized by the parameter \( \alpha \), quintessential field characterized by the parameters \(\mathrm{N}, w\), the non-commutative geometry effects characterized by the parameter \( a \), and the BH mass \( M \)). Moreover, the multipole number $\ell \geq 0$ also alter this potential.

In Figure \ref{fig:12}, we illustrate the scalar perturbative potential as a function of $r$ for different values of the string parameter \(\alpha\), NC geometry parameter \(\lambda\), and the normalization constant \(\mathrm{N}\) associated with the quintessence field. Panels (a) and (c) show that the potential decreases as \(\alpha\) and \(\mathrm{N}\) increase, respectively. Conversely, panel (b) demonstrates that the potential increasing values of the NC geometry parameter \(\lambda\).

\section{Conclusions}\label{sec:5}

In this work, we presented a comprehensive investigation of geodesic motion, thermodynamics, and scalar field perturbations in an AdS BH space-time that incorporated the effects of a cloud of strings and a quintessence-like fluid within a NC geometry background. The resulting space-time deviated from the classical Schwarzschild solution due to the presence of three key parameters: the string cloud parameter \( \alpha \), the quintessence-like fluid parameters \( (\mathrm{N}, w) \), and the NC geometry parameter \( \lambda \).

First, we conducted a systematic analysis of the geodesic motion of massless and massive particles in the selected BH solution. We derived the effective potential governing null geodesics and examined optical properties such as photon trajectories, the effective radial force on photons, the photon sphere radius, and the BH shadow size. We demonstrated how the geometric parameters-including the string cloud parameter \( \alpha \), the NC parameter \( \lambda \), the quintessence normalization constant \( \mathrm{N} \), the equation of state parameter \( w \), the BH mass \( M \), and the cosmological constant \( \Lambda \)-collectively influenced these optical characteristics. To illustrate these effects, we presented several figures showing how variations in the parameters altered the physical observables. In addition, we studied the dynamics of test particles by deriving expressions for the specific energy and angular momentum of particles in circular orbits. We analyzed the innermost stable circular orbit and have shown that the aforementioned geometrical parameters significantly influenced particle dynamics and shifted the ISCO radius. Numerical results were provided to illustrate how changes in these parameters affected the ISCO radius.

Next, we explored the thermodynamic properties of the BH space-time. We derived key thermodynamic quantities, including the Hawking temperature \( T_H \), entropy \( S \), specific heat capacity \( C_p \), and Gibbs free energy \( G \). Our results have shown that the NC parameter \( \lambda \), string cloud parameter \( \alpha \), and quintessence parameters \( (\mathrm{N}, w) \) significantly altered the BH’s thermodynamic behavior. In particular, we identified critical points and phase transitions characterized by divergences in the heat capacity and non-monotonic behavior in the Gibbs free energy. The NC geometry played a regularizing role at short distances, while the string cloud and quintessence-like fluid controlled global thermodynamic stability. Our findings extended previous results in NC geometry and vacuum AdS scenarios and provided new insights into the rich phase structure of BH space-times. We observed that the presence of string clouds increased the Hawking temperature, while the NC geometry smoothed singular behaviors in the ultraviolet regime. The quintessence parameters \( \mathrm{N} \) and \( w \) introduced significant modifications to thermal stability and global phase structure, evident from shifts in heat capacity divergence and the emergence of swallowtail-like profiles in the Gibbs free energy-signatures of first- and second-order phase transitions. Local minima in temperature and heat capacity, along with the behavior of Gibbs free energy, pointed toward possible small-to-large BH transitions analogous to the Hawking-Page phenomenon. These effects were illustrated in figures showing the variation of thermodynamic quantities as functions of the horizon radius, with parameters \( \alpha \), \( \lambda \), and \( \mathrm{N} \) varied at a fixed equation of state parameter \( w = -2/3 \).

Finally, we investigated scalar (spin-0) field perturbations in the selected BH background. Starting from the massless Klein–Gordon equation, we applied a suitable ansatz to derive the radial perturbation equation in Schrödinger-like form. The resulting effective potential was computed explicitly and shown to be significantly influenced by the parameters \( \alpha \), \( \lambda \), \( \mathrm{N} \), and \( w \), modifying the behavior relative to standard BH space-times. Scalar perturbations are well known to be useful in testing linear stability; if such perturbations decay over time, the BH is deemed stable under scalar field fluctuations. The BH’s response to these perturbations is governed by quasinormal modes (QNMs), which are damped complex-frequency oscillations depending solely on the BH’s properties and geometry. 

As directions for future research, we proposed extending the model to rotating or higher-dimensional (\( \mathcal{D} \)) BH solutions, exploring fermionic field perturbations, and analyzing the influence of electromagnetic and axionic fields. Further studies involving QNM spectra, greybody factors, and scattering cross-sections may provide valuable observational signatures, potentially allowing for constraints on the NC parameter \( \lambda \), string cloud parameter \( \alpha \), and quintessence parameters \( \mathrm{N} \) and \( w \). Additional matter fields may also be incorporated to further explore the quantum gravitational landscape.

{\small
\section*{Acknowledgments}

F.A. acknowledges the Inter University Centre for Astronomy and Astrophysics (IUCAA), Pune, India for granting visiting associateship.

\section*{Data Availability Statement}

This manuscript has no associated data.

\section*{Conflict of Interests}

Author declare (s) no conflict of interest.
}


\begin{thebibliography}{}

{\small

\bibitem{AZ1} R. M. Wald, {\tt General relativity}, University of Chicago press (2010).

\bibitem{AZ2} M. Harwit, {\tt Astrophysical concepts}, Springer Science (2006).

\bibitem{AZ3} J. Weber, Phys. Rev., {\bf 117}(1), 306 (1960).

\bibitem{AZ4} S. A. Hayward, Phys. Rev. {\bf D 49}(12), 6467 (1994).

\bibitem{AZ5} N. Arkani-Hamed, D. P. Finkbeiner, T. R. Slatyer, and N. Weiner,  Phys. Rev. {\bf D 79}(1), 015014 (2009).

\bibitem{AZ6} E. J. Copeland, M. Sami, and S. Tsujikawa,  Int. J. Mod. Phys. {\bf D 15}(11), 1753-1935 (2006).

\bibitem{AZ7} L. Modesto, Int. J. Theor. Phys. {\bf 49}, 1649–1683 (2010).

\bibitem{AZ8} A. Ashtekar, and J. Lewandowski, Class. Quantum Gravity {\bf 21}, R53 (2004).

\bibitem{AZ9} J. Alfaro, H. A. Morales-Tecotl, and L. F. Urrutia, Phys. Rev. {\bf D 65}(10), 103509 (2002).

\bibitem{AZ10}L. E. Ibanez, and A. M. Uranga, {\tt String theory and particle physics: An introduction to string phenomenology}, Cambridge University Press (2012).

\bibitem{AZ11} R. Gambini, and J. Pullin, Phys. Rev. Lett. {\bf 110}, 211301 (2013).

\bibitem{AZ12} S.G. Ghosh, and R.K. Walia,  Ann. Phys. (NY) {\bf 434}, 168619 (2021).

\bibitem{AZ13} M. Afrin, and S.G. Ghosh, Astrophys. J. {\bf 932}, 51 (2022).

\bibitem{AZ14} J. Olmedo, S. Saini, and P. Singh,  Class. Quantum Gravity {\bf 34}, 225011 (2017).

\bibitem{AZ15} A. Barrau, and C. Rovelli, Phys. Lett. {\bf B 739}, 405–409 (2014).

\bibitem{AZ16} E. Battista, Phys. Rev. {\bf D 109}(2), 026004 (2024).

\bibitem{AZ17} K. Akiyama, {\it et al.},  Astrophys. J. Lett. {\bf 875}, L1 (2019).

\bibitem{AZ18} K. Akiyama, {\it et al.},, Astrophys. J. Lett. {\bf 930}, L17 (2022).

\bibitem{AZ19} S.E. Gralla, and A. Lupsasca, Phys. Rev. {\bf D 102}, 124003 (2020).

\bibitem{AZ20} M.D. Johnson, et al., Sci. Adv. {\bf 6}, eaaz1310 (2020).

\bibitem{AZ21} C. Liu, T. Zhu, Q. Wu, K. Jusufi, M. Jamil, M. Azreg-Aïnou, and A. Wang, Phys. Rev. {\bf D 101}, 084001 (2020).

\bibitem{AZ22} O.Y. Tsupko, Phys. Rev. {\bf D 106}, 064033 (2022).

\bibitem{AZ23} S.E. Gralla, and A. Lupsasca, Phys. Rev. {\bf D 101}, 044031 (2020).

\bibitem{AZ24} T. Zhu, Q. Wu, M. Jamil, and K. Jusufi,  Phys. Rev. {\bf D 100}, 044055 (2019).

\bibitem{AZ25} C. Liu, T. Zhu, and Q. Wu, Chin. Phys. {\bf C 45}, 015105 (2021).

\bibitem{NC1} P. Nicolini, Int. J. Mod. Phys. A \textbf{24}, 1229 (2009). 

\bibitem{NC2} R.J. Szabo, Class. Quant. Grav. \textbf{23}, R199 (2006).

\bibitem{NC3} H.S. Snyder, Phys. Rev. \textbf{71}, 38 (1947). 

\bibitem{NC4} A. Smailagic, E. Spallucci, J. Phys. A \textbf{36}, L467 (2003).

\bibitem{NC5} A. Smailagic, E. Spallucci, J. Phys. A \textbf{36}, L517 (2003). 

\bibitem{NC7} P. Nicolini, A. Smailagic, E. Spallucci, Phys. Lett. B \textbf{632}, 547 (2006). 

\bibitem{NC6} L. Susskind, Phys. Rev. Lett. \textbf{71}, 2367 (1993).

\bibitem{NC8} N. Seiberg, E. Witten, JHEP \textbf{09}, 032 (1999). 

\bibitem{NC11} S. Ansoldi, P. Nicolini, A. Smailagic, and E. Spallucci, Phys. Lett. B \textbf{645}, 261 (2007).  

\bibitem{NC13} L. Modesto, P. Nicolini, Phys. Rev. D \textbf{82}, 104035 (2010). 

\bibitem{NC12} P. Nicolini, G. Torrieri, JHEP \textbf{08}, 097 (2011).

\bibitem{NC14} K. Nozari, S.H. Mehdipour, JHEP \textbf{03}, 061 (2009). 

\bibitem{NC15} T.G. Rizzo, JHEP \textbf{09}, 021 (2006).

\bibitem{NC16} E. Spallucci et al., Phys. Lett. B \textbf{670}, 449 (2009). 

\bibitem{NC17} K. Nozari, S.H. Mehdipour, Commun. Theor. Phys. \textbf{53}, 503 (2010). 

\bibitem{NC9} K. Nozari, S.H. Mehdipour, Class. Quant. Grav. \textbf{25}, 175015 (2008).

\bibitem{NC18} R. Banerjee et al., Phys. Rev. D \textbf{77}, 124035 (2008).

\bibitem{NC19} Y.S. Myung et al., JHEP \textbf{22}, 012 (2007). 

\bibitem{NC20} K. Nozari, B. Fazlpour, Acta Phys. Polon. B \textbf{39}, 1363 (2008) 

\bibitem{NC21} Y. Miao, Z. Xu, Eur. Phys. J. C \textbf{76}, 217 (2016). 

\bibitem{NC22} Z. Yan, C. Wu, and W. Guo, Nucl. Phys. B \textbf{961}, 115217 (2020). 

\bibitem{NC23} Z. Yan, C. Wu, and W. Guo, Nucl. Phys. B \textbf{973}, 115595 (2021). 

\bibitem{NC24} K. A. Bronnikov, R. A. Konoplya, and A. Zhidenko, Phys. Rev. D \textbf{86}, 024028 (2012). 

\bibitem{NC25} R. A. Konoplya, A. F. Zinhailo, J. Kunz, Z. Stuchlik, and A. Zhidenko, JCAP \textbf{10}, 091 (2022). 

\bibitem{NC26} J. A. H. Futterman, F. A. Handler, and R. A. Matzner, \textit{Scattering from Black Holes} (Cambridge University Press, 1988). 

\bibitem{NC27} K. Glampedakis, and N. Andersson, Class. Quant. Grav. \textbf{18}, 1939 (2001).

\bibitem{NC28} C. Doran, A. Lasenby, S. Dolan, and I. Hinder, Phys. Rev. D \textbf{71}, 124020 (2005).

\bibitem{NC29} S. Dolan, C. Doran, and A. Lasenby, Phys. Rev. D \textbf{74}, 064005 (2006). 

\bibitem{NC30} L.C.B. Crispino, E. S. Oliveira, A. Higuchi, and G. A. Matsas, Phys. Rev. D \textbf{75}, 104012 (2007). 

\bibitem{NC31} S.R. Dolan, Class. Quant. Grav. \textbf{25}, 235002 (2008).

\bibitem{NC32} E. Jung, S. H. Kim, and D. Park, Phys. Lett. B \textbf{602}, 105 (2004). 

\bibitem{NC33} C. Doran, A. Lasenby, S. Dolan, and I. Hinder, Phys. Rev. D \textbf{71}, 124020 (2005). 

\bibitem{NC34} S. Dolan, C. Doran, and A. Lasenby, Phys. Rev. D \textbf{74}, 064005 (2006). 

\bibitem{NC35} J. Castineiras, L. C. Crispino, and D. P. M. Filho, Phys. Rev. D \textbf{75}, 024012 (2007). 

\bibitem{NC36} C.L. Benone, E. S. de Oliveira, S. R. Dolan, and L. C. Crispino, Phys. Rev. D \textbf{95}, 044035 (2017). 

\bibitem{NC37} J. Chen, H. Liao, Y. Wang, and T Chen, Eur. Phys. J. C \textbf{73}, 2395 (2013). 

\bibitem{NC38} L.C.B. Crispino, S. R. Dolan, and E. S. Oliveira, Phys. Rev. D \textbf{79}, 064022 (2009). 

\bibitem{NC39} H. Huang, P. Liao, J. Chen, and Y. Wang, J. Grav. \textbf{2014}, 231727 (2014).

\bibitem{NC40} M. Anacleto, F. A. Brito, J. A. V. Campos, and E. Passos, Phys. Lett. B \textbf{803}, 135334 (2020). 

\bibitem{NC41} M.A. Anacleto, F.A. Brito, and E.  Passos, Phys. Lett. B \textbf{743}, 184 (2015). 

\bibitem{NC42} F. Moura, JHEP \textbf{09}, 038 (2013). 

\bibitem{RBW} R.-B. Wang, S.-J. Ma, L. You, J. -B. Deng and X.-R. Hu, Chin. Phys. {\bf  C 49}, 065101 (2025).

\bibitem{JMT} J. M. Toledo and V. B. Bezerra, Eur. Phys. J. C {\bf 79}, 110 (2019).

\bibitem{FA1} F. Ahmed, A. Al-Badawi, I. Sakallı and A. Bouzenada, Nucl. Phys. {\bf B 1011}, 116806 (2025) 

\bibitem{FA2} A. Al-Badawi and F. Ahmed, Chin. J. Phys. {\bf 94}, 185 (2025). 

\bibitem{FA3} F. Ahmed, J. Goswami and A. Bouzenada, Eur. Phys. J C {\bf 85}, 110 (2025).

\bibitem{FA4} F. Ahmed, A. Al-Badawi, and I Sakalli, Phys. Dark Univ. {\bf 48}, 101925 (2025). 

\bibitem{FA5} F. Ahmed, A. Al-Badawi, I. Sakallı and S. Kanzi, Phys. Dark Univ. {\bf 48}, 101907 (2025). 

\bibitem{FA6} F. Ahmed, A. Al-Badawi, and I. Sakallı, Nucl. Phys.{\bf B 1017}, 116951 (2025).

\bibitem{FA7} A. Al-Badawi, F. Ahmed, and İ. Sakallı, Eur. Phys. J. C {\bf 85}, 660 (2025).

\bibitem{FA8} F. Ahmed, A. Al-Badawi, and İ. Sakallı, Eur. Phys. J. C {\bf 85}, 668 (2025). 

\bibitem{FA9} A. Al-Badawi, F. Ahmed, I. Sakallı, S. Shaymatov, Chin. J. Phys. {\bf 96}, 770 (2025). 

\bibitem{FA10} F. Ahmed, A. Al-Badawi, and İ. Sakallı, Eur. Phys. J. C {\bf 85}, 554 (2025). 

\bibitem{FA11} A. Al-Badawi, F. Ahmed, and I. Sakallı,  Nucl. Phys. B {\bf 1017}, 116961 (2025).

\bibitem{FA12} F. Ahmed, A. Al-Badawi, and İ. Sakallı, Phys. Dark Univ. {\bf 49}, 101988 (2025).

}

\end{thebibliography}
\end{document}